\documentclass[twocolumn,rmp,showpacs,floatfix]{revtex4}
\usepackage{graphics}
\usepackage{graphicx}

\begin{document}

{\textbf Reviews of Modern Physics (in press)}

\vspace{0.6cm}

%%%\leftline{{\bf Journal Ref:} {\large{{\it Reviews of Modern Physics}}} (in press)}

%%\vspace{1cm}

\title{Quantum Annealing and Analog Quantum Computation}
\author{Arnab Das}
\email{arnab.das@saha.ac.in}
\author{Bikas K. Chakrabarti}
\email{bikask.chakrabarti@saha.ac.in}
\affiliation{Theoretical Condensed Matter Physics Division and \\
Centre for Applied Mathematics and Computational Science,
Saha Institute of Nuclear
Physics, 1/AF, Bidhannagar, Kolkata-700064, India.}

\begin{abstract}
We review here the recent success in quantum annealing, i.e., optimization 
of the cost or energy functions of complex systems 
utilizing quantum fluctuations. 
The concept is introduced in successive steps 
through the studies of  
mapping of such 
computationally hard problems to the classical spin glass problems.
The quantum spin glass problems arise with the introduction
of quantum fluctuations, and the annealing behavior of the systems as
these fluctuations are reduced slowly to zero. This provides a general
framework for realizing analog quantum computation.
\end{abstract}


\maketitle

\tableofcontents

\section{Introduction}

\vspace{0.65cm}

\noindent
Utilization of
quantum mechanical tunneling through classically localized states in 
annealing of glasses
has opened up a new paradigm for solving hard optimization problems through
adiabatic reduction of quantum fluctuations. 
This will be introduced and reviewed here. 
%We introduce the
%idea in details through successive steps.

%%%%%%%%%%%%%%%%%%%%%%%%% BKC-ADD: BEGINS %%%%%%%%%%%%%%%%%%%%%%%%%%%%%%%%%%%%%%
Consider the example of a ferromagnet consisting of $N$ tiny interacting
magnetic elements; the spins. For a macroscopic sample $N$ is very large;
of the order of Avogadro number. 
Assume that each spin can be in any of the
two simple states: up or down. Also, the pairwise interactions between the
spins are such that the energy of interaction (potential energy or PE) 
between any pair of spins is negative (smaller) if both the spins in the
pair are in the same state and is positive (higher) if their states differ.
As such, the collective energy of the $N$-spin system
(given by the Hamiltonian ${\mathcal H}$) is minimum when
all the spins are aligned in the same direction; all up or all down, giving
the full order. 
We call these two minimum energy configurations 
the ground states.
The rest of the $2^{N}$ configurations are called 
excited states. 
The plot of the interaction energy of the whole system 
with respect to the 
configurations is called the 
potential energy-configuration landscape, or simply, the
potential energy landscape (PEL).
%To clarify further, one may think, for an example,
% of a particle moving in one dimension, say, along the 
%$x$-axis. The PEL for the particle will simply be the
%plot of its potential energy $V(x)$
%with respect to its position $x$.
For a ferromagnet, this landscape has a
smooth double-valley structure (two mirror-symmetric 
valleys with the two 
degenerate ground states, 
all up and all down, at their respective bottoms). 
At zero-temperature the equilibrium state is
the state of minimum potential energy, and the system stably
resides at the bottom of any of the two valleys. 
At finite temperature,
the thermal fluctuations allow the system to visit higher energy 
configurations with some finite 
probability (given by the Boltzmann factor) 
and thus the system spends time in other part
of the PEL also.  
The probability that a system
is found in a particular macroscopic state 
depends not only on the energy of the state (as at zero temperature), 
but also on its entropy. The thermodynamic 
equilibrium state corresponds to the
minimum of a thermodynamic potential called free energy $F$
given by the 
%instead of minimizing the potential energy itself. For a system at a constant
%temperature, the free energy of a state is given by the 
difference between the energy of the state and the product of its entropy and the 
temperature. 
%The plot of the free energy with respect to 
%the order parameter (for a ferromagnet it is the plot of $F(m)$ against $m$)
%is called the free energy landscape. 
At zero-temperature, the minima of the free energy
coincides with the minimum for energy and one gets the highest order (magnetization). 
As the temperature is increased, the contribution of entropy gets magnified and the
minimum of the free energy is shifted more and more towards the
states with lower and lower 
order or magnetization, until at (and beyond) 
some transition temperature $T_{c}$,
the order disappears completely.
For antiferromagnetic systems, the spin-spin pair 
interactions are such that it is lower if the spins in the pair
are in opposite states, and higher if their states are same. For
antiferromagnets one can still define a sub-lattice order or magnetization
and the PEL still has the double-well structure like in a ferromagnet for
short range interactions. The free energy and the order-disorder transition
also shows identical behavior as observed for a ferromagnet.

In spin glasses, where different 
spin-spin interactions are randomly ferromagnetic
or antiferromagnetic and frozen in time (quenched disorder), the
PEL becomes extremely rugged; various local and global minima 
trapped between potential energy barriers
appears in the PEL. The ruggedness and the degeneracies in the
minima comes from the effect of frustration or competing interactions
between the spins; none of the spin states on a cluster or a placket
is able to satisfy all the interactions in the cluster. The locally 
optimal state for the spins in the cluster is therefore 
degenerate and frustrated. 

Similar situations occur for multivariate optimization problems
like the traveling salesman problem (TSP). Here a salesman has to
visit $N$ cities placed randomly on a plane (country). Of the $N!/N$
distinct tours passing through each city once, only few
corresponds to the minimum (ground state) travel distance or travel cost.  
The rests corresponds to the higher costs (excited states). The cost
function, when plotted against different tour configurations, gives
a similar rugged landscape, equivalent to the PEL of a spin glass
(henceforth we will use PEL to mean cost-configuration landscapes
also). 
  
\begin{figure}
\resizebox{5.0cm}{!}{\includegraphics{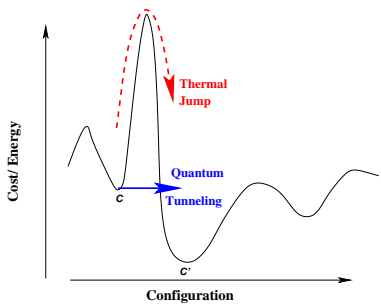}}
\caption{\footnotesize{While optimizing the cost function of a
computationally hard problem (like the  
ground state energy of a
spin glass or the minimum travel distance for a
traveling salesman problem), one has to get out of a shallower
local minimum like the configuration C (spin configuration or travel route), 
to reach a deeper minimum C$^{\prime}$. 
This requires jumps or tunneling like fluctuations in the dynamics.
Classically one has to jump over the energy or the cost barriers 
 separating them,
while quantum mechanically one can tunnel through the same.
If the barrier is high enough, thermal jump becomes very difficult. 
However, if 
the barrier
is narrow enough, quantum tunneling often becomes quite easy.}}
\label{QA-SA}
\end{figure}

Obviously, an exhaustive search for the global minimum of a rugged
PEL requires an exponentially large (or higher) number (in $N$) of
searches ($2^{N}$ or $N!$ order of searches for an $N$-spin spin glass 
and an $N$-city TSP respectively). The computational effort or time for
such searches are therefore generally not bounded by any polynomial 
in the problem size $N$. 
Alternatively, a gradual energy or cost dissipative
dynamics (annealing), with Boltzmann like thermal fluctuations
or some noise factor (in order to get out of the the local minima)
in the PEL may help reaching deep enough minima  
more easily. This simulated thermal annealing scheme is
considered to be a very successful technique now. However,
such a technique often fails if the barrier heights between the
minima diverges (or become very high),
as in case of a spin glass, due to 
frequent trapping of the system in such local minima
(glassy behavior). 
In case such barriers are
very narrow, quantum-mechanical fluctuations 
(fluctuations in a quantum observable
due to its non-commutativity with 
the Hamiltonian of the system) 
can help tunneling through
them, thereby leading to 
successful quantum annealing (see. Fig. \ref{QA-SA}).    

We introduce here these ideas in details through successive
steps.

%%%%%%%%%%%%%%%%%%%%%%%%% BKC-ADD: ENDS  %%%%%%%%%%%%%%%%%%%%%%%%%%%%%%%%%%%%%%%

\noindent
 a) The physics of classical spin glasses has 
already contributed enormously to our knowledge of the landscape structure
of the energy or the thermodynamic potentials 
and that of the
unusually slow (glassy) dynamics of many-body
systems in the presence of frustration and disorder. Mapping of 
computationally
hard problems, like the traveling salesman problem etc,
to classical spin glass models also helped understanding their
complexity.

\noindent
 b) The ground (and some low-lying) state structures of frustrated random
systems in the presence of quantum fluctuations 
have also been studied in the context of
quantum spin glasses. 
It has been shown that
because of the possibility of 
%%%%% corr-ref-2-3 %%%%%%%%%
tunneling through the barriers
in the potential energy landscape,
quantum fluctuations can 
help the dynamics to
be `more ergodic' than
the dynamics induced by
the classical fluctuations and
thus help exploring 
the landscape much better.
Ergodicity  here means the loss of the
memory of the initial state in course of evolution (weak ergodicity)
and the convergence to a stationary distribution irrespective of the
the initial state (strong ergodicity). 
The nature of these quantum phase transitions
in such systems have also been extensively studied. These
studies (Sec. IID) endow one with the knowledge of the
phase diagram and the location of the quantum 
critical point or the phase boundary, 
which is crucial for choosing the proper quantum kinetic terms and 
the annealing path (Sec. IIIA and Sec. IIIB).

\noindent
 c) The most natural connection between the paradigm of
classical spin glasses and hard optimization problems comes 
through a widely used and well established optimization
technique, namely, simulated annealing algorithm as 
discussed earlier. 
The possibility of quantum tunneling through 
classically impenetrable barriers, as indicated
from the studies of quantum spin glasses, naturally suggests
an elegant and often more effective alternative to simulated annealing.
%This novel concept, known as quantum annealing, employs
%quantum fluctuations to anneal a glassy system to its ground state
%instead of thermal fluctuations (see Fig. \ref{QA-SA}). 

In quantum annealing, one has a  
classical Hamiltonian 
(or a multivariate cost function viewed as the same) to be optimized,
to which, one adds a (non-commuting) quantum kinetic term and reduce it
from a very high initial value to zero eventually. 
%This effectively tunes the Planck's
%constant to zero to reach the classically optimized (minimum cost) state.
This reduction, when
done completely adiabatically, assures 
reaching of the ground state of the classical 
glass at the end, assuming that there is no
crossing of energy levels with the ground state in the course of
evolution, and provided that the starting state was 
the ground state of
the initial Hamiltonian. To start with, the tunneling field is much
higher
than the interaction term, so the ground state 
(a uniform superposition of all classical configurations) is
trivially realizable. 
Simulations clearly demonstrate %%%%%% corr-r1-1 %%%%%%%%%%
that quantum annealing can occasionally help 
reaching the ground state of a complex
glassy system much faster than 
could be done using thermal annealing 
(discussed later in Sec. III). 
An experiment comparing the classical
and quantum annealing for a spin glass
also shows that the relaxations in course of quantum annealing are
often much faster than those during the corresponding 
classical annealing, 
%(Brooke et al 1999) 
as discussed in Sec. IIID. What makes quantum annealing
fundamentally different from the classical annealing, is the
non-local nature (Sec. III) and its higher 
tunneling ability (Secs. IID \& IIIC).

%difference in nature of fluctuations (quantum and classical) 
%in the two cases. Unlike the classical fluctuations (which is local)
% quantum fluctuations allow the system to sense the whole PEL
%simultaneously through an extended wave function. 
%Quantum fluctuations also helps tunneling through 
%classically impenetrable high energy barriers 
%if they are narrow enough (see Sec. IIIB and Secs. IID \& IIIC).
 
%%%%%%%%%%%%%%%%% corr-r1-1 %%%%%%%%%%%%%%%%%%%%%%%%%%%%%%%%%%%%
%Thus 
Quantum annealing thus permits a realization 
of analog quantum computation, 
which is an independent and powerful 
complement to digital quantum computation, 
where discrete unitary transformations
are implemented through quantum logic gates. 

%\section{Classical Spin Glasses}
\section{Optimization and Annealing}

\subsection{Combinatorial Optimization Problems}
\noindent
The occurrence of
multivariate optimization problems is ubiquitous
in our life, wherever one has to 
choose the best bargain from a host of available options
that depend on many independent factors. 
In many cases, 
such a task can be cast as a problem of minimizing a given cost 
or energy function
$\mathcal{H}(S_1, S_2,...S_N)$ with respect to 
$N$ variables $S_1, S_2,...S_N$
(sometimes subject to some constraints). 
The task is to find a set
of values
for these variables (a configuration) 
for which the function $\mathcal{H}(\{S_i\})$ has the 
minimum value (cf. Fig. \ref{QA-SA}).
In many important 
optimization problems, the set
 of feasible configurations 
from which an optimum is to be chosen
 is a finite set (for finite $N$). 
In such a case, we say that the problem is
 combinatorial in nature. 
If the variables $S_i$ are discrete and each takes
up a finite number of values, then the problem is clearly a 
combinatorial one. Moreover, 
 certain problems with continuous variables 
(like linear programming problem) 
can also be reduced to combinatorial
problems (Papadimitriou et al 1998). 
Here we focus on this type of optimization problem,
and assume that we have to minimize
 $\mathcal{H}(\{S_i\})$ with respect
to the discrete set of the variables $S_i$. 

An optimization problem is said to belong to the
class P (P for Polynomial), 
if it can be solved in polynomial time (i.e., the
evaluation time goes like some polynomial in $N$) 
using polynomially (in $N$, again) bound resources
(computer space, processors etc).
% deterministic Turing machine or by any computational
%model polynomially equivalent to it (Garey and Johnson 1979).   
Existence of such a polynomial bound
on the evaluation time is somehow interpreted 
as the ``easiness'' of the
problem. 
However, many important optimization 
problems seem to fall outside
this class, like, the famous traveling salesman problem (see Sec. IIC.2).

There is another important class of problems which can be solved
in polynomial time by non-deterministic machines. 
This class is the famous
 NP (Non-deterministic Polynomial) class
(Garey and Johnson 1979). P is included completely
in the NP class, since a deterministic Turing machine is just a special
case of non-deterministic Turing machines. Unlike a deterministic machine,
which takes a specific step deterministically at each instant 
(and hence follows a single computational path), 
a non-deterministic machine has a host of different `allowed' 
steps at its disposal at every instant. At each instant
it explores all the `allowed' steps and if any of them leads to the
goal, the job is considered to be done. Thus it explores in parallel
many paths (whose number goes roughly exponentially with time) and
checks if any one of them reaches the goal. 
   
Among the NP problems, there
are certain problems (known as NP-complete problems) which are such that any 
NP problem
can be ``reduced'' to them using a polynomial algorithm. The famous
3-SAT problem (see Sec. IIIA.3) is a representative of the class.  
This roughly means that if one has a 
routine to solve an
NP-complete problem of size $N$ then using that routine one can 
 solve any NP problem at the cost of an extra overhead in time that
goes only polynomially with $N$.
The problems in this class are considered to be hard, since so far
no one can simulate a general nondeterministic machine by a 
deterministic Turing
machine (or any sequential 
computer with polynomially bound resources) 
without an exponential growth of execution time. 
In fact
it is largely believed (though not proved yet) 
that it is impossible to
do so (i.e., P$\ne$NP) in principle. 
However, assuming this to be true, one can show that
there are indeed problems in NP class that are neither NP-complete nor P 
(Garey and Johnson 1979). 

\subsection{Statistical Mechanics of the Optimization Problems and Thermal 
Annealing}
\noindent
There are some excellent deterministic algorithms
for solving certain optimization problems exactly 
(Papadimitriou and Steiglitz 1998, Hartmann and Rieger, 2002). 
These algorithms are, however, quite small in number and are
strictly problem specific. For NP or harder problems, only approximate
results can be found using these algorithms in polynomial time. These 
approximate algorithms too are also strictly problem specific in
the sense that 
if one can solve a certain NP-complete
problem up to a certain approximation using some polynomial algorithm, then
that does not ensure that one can solve all other NP problems using 
it up to the said approximation in
polynomial time.    

Exact algorithms being scarce, one has to go for heuristics algorithms,
which are algorithms based on certain intuitive moves,  
without any   guarantee on either the accuracy or the run time for the
worst case instance. However, these algorithms are generally easy to
formulate and are quite effective in solving most instances of a the 
intended problems.
A general approach towards formulating such approximate heuristics 
may be based on stochastic (randomized) iterative
improvements. The most preliminary one in this family is the local minimization
algorithm. In this algorithm 
one starts with a random configuration $C_0$ and makes some
local changes in the configuration following some
 prescription (stochastic or deterministic) 
to generate a new configuration $C_1$
and calculates the corresponding change in the cost. 
If the cost is lowered by the change,
then the new configuration $C_1$ is adopted. Otherwise the old
configuration is retained. Then in the
next step a new local change is attempted again, and so on. 
This reduces the cost
steadily until a configuration is reached which minimizes the cost locally.
This means that no further lowering of cost is possible by changing this
configuration using any of the prescribed local moves. The algorithm 
essentially stops there. But generally 
in most optimization problems (such as in spin glasses),  
there occur
many local minima in the cost-configuration landscape and
they are mostly far above
the global minimum (see Fig. \ref{QA-SA}). It is
likely that the algorithm 
therefore gets stuck in one of them and ends up with a
poor approximation. One can then start afresh with some new initial 
configuration and end up with another local minimum. After repeating this for
several times, each time with a new initial configuration, one may choose the
best result from them. But much better idea would be to get somehow out
of shallow local minima. One can 
introduce some fluctuations or noise in the process
 so that the movement is not always towards lower
energy configurations, but there is 
also a finite probability to go to 
 higher
energy configurations (the higher the final energy, 
the lower the probability to move to that), 
and consequently there appear chances 
to get out of the shallow local minima. Initially, 
 strong fluctuations are adopted (i.e., the probability to go to higher
energy configurations is relatively high) and slowly 
fluctuations are reduced until 
finally they are tuned off completely. In the mean time the system gets a
fair opportunity to explore the landscape more exhaustively
and settle into a reasonably deep cost or energy minimum.
Kirkpartick et al (1983) suggested an elegant way: 
A fluctuation
is implemented by introducing an ``artificial'' 
temperature $T$ into the problem 
such that
the transition probability from a configuration $C_i$ to
a configuration $C_f$ is given by $\min{\{1,\exp-{[\Delta_{if}/T]}\}}$,
where $\Delta_{if} = E_{f} - E_{i}$, with $E_k$ denoting the cost
or energy 
of the configuration $C_k$. A corresponding Monte Carlo dynamics is defined,
say, based on detailed balance, and the thermal relaxation of the system is 
simulated. In course of simulation, the noise factor $T$ 
is reduced slowly from a very high initial value to zero
following some annealing schedule. At the end of the simulation one is
expected to end up with a configuration whose cost is a reasonable 
approximation of the globally minimum one. 
If the temperature is decreased 
slow enough, say,  
\begin{equation}
T(t) \ge N/\log{t},
\label{SA-conv}
\end{equation}
\noindent
where
 $t$
denotes the cooling time and $N$ the system size, 
then the global minimum 
is attained with certainty in the limit
$t \rightarrow \infty$ (Geman and Geman 1984). 
Even within a finite time and with a faster
cooling rate, one can achieve a reasonably good approximation (a crystal with
only few defects) in practice. 
This 
simulated annealing method is now being used extensively by engineers for
devising real-life optimization algorithms. We will refer to this as the
classical annealing (CA), to distinguish it from quantum annealing (QA)
which employs quantum fluctuations. 
It is important to note that though in this
type of stochastic algorithms the system has many different steps with 
their respective probabilities at its disposal, it finally takes up a single 
one, say by tossing coins, and thus finally follow a single 
(stochastically selected) path. 
Hence it is not equivalent to a 
non-deterministic machine, where all the allowed paths are
checked in parallel at every time-step.     
 
%\subsection{Statistical Mechanics of Optimization Problems and 
%Thermal Heuristics}  
%\noindent
As has been mentioned already, many combinatorial optimization problems
can be cast into the problem of finding the ground state of some
classical
(spin glass like) Hamiltonian $H(\{S_i\})$. One can therefore  
 analyze the problem by using statistical mechanics so as to 
%%in order to be able 
to apply physical techniques like simulated annealing.
If one naively takes the number of variables $N$ as the
size, then the entropy and the energy are often found to scale differently
with $N$ and the application of 
standard thermodynamic arguments become difficult. One needs to scale
temperature and some other quantities properly with $N$ so that
one can talk in terms of the concepts like free energy minimization etc. 
Moreover, the constraints present in the problems are often very difficult
to take into account. 
%The constraints are generaly dealt with either by
%assigning heavy penalties to the forbidden configurations (Mezard et al 1987)
%or by restricting the calculation (say, the Monte Carlo moves, when 
%done numerically) within the
%space that excludes the forbidden ones, both of which can complicate the
%task considerably.
%Two different approaches are
%adopted generally to deal with the constraints. One is the softer method,
%where all the configurations are allowed in the sampling, but 
%heavy penalties are associated with those
%which are to be excluded to satisfy the constraints (Mezard et al 1987). 
%The other
%method is to restrict the calculation in the specific subspace of the
%configuration space which satisfies the constraints. Thus,
%for example,  %%%% corr-r1-5 %%%%%%%  
%in Monte Carlo
%simulation of such a system, only those specific moves are allowed 
%which satisfy the constraints.

%%%%%%%%%%%%%%%%%%%%%%%%%%%%% 

\subsection{Spin Glass and Optimization}
\subsubsection{Finding The Ground States of the Classical Spin Glasses}
\noindent
As has been mentioned already,
the difficulty faced by a physically
motivated optimization 
heuristic (one that follows
physical relaxation dynamics, classical or quantum,
to search for the solution), in finding the solution of a hard
optimization problem, is similar to that faced
by a glassy system in reaching its ground state. In fact,
finding the ground states of the spin 
glasses is an important
class of combinatorial optimization problem, 
which includes an NP-complete problem (Barahona 1982),
and many other
apparently different ones (like  the traveling salesman problem) 
can be recast in this form. Hence here we discuss
very briefly the nature of the spin glass 
phase and the difficulty in reaching its ground state.
  
%Spin glasses are random magnetic systems which show random
% freezing of the spins or magnetic moments below a certain 
%noise level (here parametrized by the temperature
% $T_c$). The main features 
% essential for such freezing are the presence of
%quenched randomness and frustration 
%(which means that there are such competing interactions that all of them
% cannot be satisfied by any spin configuration)
%in the spin-spin interactions. The presence of random 
%ferromagnetic and antiferromagnetic bonds (interactions) 
%in a spin system can produce such
%a situation 
The interaction energy of a typically 
 random and frustrated Ising spin glass 
(Binder and Young 1986, Nishimori 2001, Dotsenko 2001)
may be represented by a Hamiltonian of the form

\begin{equation}
\mathcal{H} = -\sum_{i>j}^N J_{ij}S_{i}S_{j}, 
\label{spg-ham}
\end{equation} 
\noindent
where $S_{i}$ denote the Ising spins and $J_{ij}$ the interactions between them.
These $J_{ij}$s here are quenched variables which
vary randomly both in sign and magnitude following some distribution
$\rho(J_{ij})$. The typical distributions are (zero mean)
Gaussian distribution of positive and negative $J_{ij}$
values
\begin{equation}
\rho(J_{ij}) = 
%%%%%%%% corr-r1-2 %%%%
%%A\exp{\left(\frac{-NJ_{ij}^2}{2J^2}\right)},  \label{Gaussian} 
A\exp{\left(\frac{-J_{ij}^2}{2J^2}\right)},  \label{Gaussian} 
\end{equation}
\noindent
and binary distribution
\begin{equation}
\rho(J_{ij}) = p\delta (J_{ij} - J) + (1 - p)\delta (J_{ij} + J),
\label{pls-mns-J}
\end{equation}
\noindent
with probability $p$ of having a $+J$ bond, and $(1-p)$ of having a
$-J$ bond. Two well studied models are the SK model 
(introduced by Sherrington and Kirkpatrick 1975) and the EA model
(introduced by Edward and Anderson 1975). In the SK model
the interactions are infinite-ranged and for the sake
of extensively (for rules of equilibrium thermodynamics to be applicable,
energy should be proportional to the volume or its equivalent that 
defines the system-size) 
one has to scale $J \sim 1/\sqrt{N}$, %%%% corr-r1-2 %%%%% 
while in the EA model, the interactions are
between the nearest neighbors only. 
For both of them, however, $\rho(J_{ij})$ is Gaussian;
$A = (N/2\pi J^{2})^{1/2}$ for normalization in the SK model.

%\subsection{Ordering}
The freezing (below $T_{c}$) is characterized by some non-zero value of the
thermal average of the magnetization 
at each site (local ordering). 
However, since the interactions are random and competing, 
the spatial average
of single-site magnetization (below $T_c$) is zero. Above $T_c$, both
the spatial and the temporal averages of the 
single-site magnetization of course
vanishes. 
A relevant order parameter for this freezing is therefore 
\begin{equation}
q = \frac{1}{N}\overline{\sum_{i}^{N} 
%\langle\langle 
\langle
S_{i}\rangle_{T}^{2}}, 
\label{EA-ord}
\end{equation}
\noindent (overhead bar denoting the average over disorders 
(the distribution of $J_{ij}$) and  
%$\langle\langle 
$\langle 
... \rangle_{T}$ denotes the thermal average).
Here $q \ne 0$ for $T < T_c$,
while $q = 0$ for $T \ge T_c$. As will be seen in the following,
the existence of a unique order parameter $q$ will indicate ergodicity.

In the spin glass phase ($T<T_{c}$), the whole free 
energy landscape gets divided (cf. Fig. \ref{QA-SA}) into
many valleys (local minima of free energy) separated by 
very high free energy barriers. Thus the system, once trapped
in a valley, remains there for a very long time.
%%%%%%%%%%%%%%%%%%%%%%%%%%%%%%%%%%%%%%%%%%%%%%%%%%%%%%%%%%%%%%%%%%%%%
%For infinite-range spin glasses,
%the barriers may become infinitely high in the thermodynamic 
%limit ($N\rightarrow\infty$; 
%see Binder and Young 1986, Nishimori 2001). The 
%valleys then get completely separated from each other, so that
%if the system initially
%be within one valley, then it remains confined there for 
%infinitely
%long time
%and the spin glass system 
%thus becomes non-ergodic. 
%Hence the observable free energy 
%(and hence local magnetization and 
%all other 
%physical quantities) for such a trapped system is given 
%by the averages over the configurations within
%that valley only, instead of the
% one over the whole configuration space. 
%Consequently, as the ergodicity is lost,
%and a distribution 
%$P(q)$ of order
%parameter emerges corresponding to the distribution of these
%thermodynamically separated frozen states. 
%%%%%%%%%%%%%%%%%%%%%%%%%%%%%%%%%%%%%%%%%%%%%%%%%%%%%%%%%%%%%%%%%%%%%%
%The valleys are thermodynamically stable. 
%and are termed as pure states. 
The spins of such a confined system
are allowed to explore only a very restricted
(and correlated) part of the 
configuration space and thus makes them 
``freeze'' to have a magnetization
that characterizes the state (valley) locally.

So far, two competing pictures continue to represent the
physics of the spin glasses. The mean-field
picture of replica symmetry-breaking
is valid for infinite-ranged spin glass systems
like the SK spin glass. In this picture, below the glass transition
temperature $T_{c}$, the barriers separating the
valleys in the free energy landscape actually diverge 
(in the limit $N\rightarrow\infty$), giving rise
to a diverging timescale for the confinement of the system in any such
valley once the system gets there somehow. This means there is
a loss of ergodicity in the thermal dynamics of the system at $T < T_{c}$. 
Thus one needs a distribution $P(q)$ of order parameters, instead of a 
single order parameter to characterize the whole landscape, 
as emerges naturally from the replica symmetry-breaking 
ansatz of Parisi (1980). To be a bit more quantitative,
%%%%%%%%%%%%%%%%%%%%%%%%%%%%%%%%%%%%%%%%%%%%%%%%%%%%%%%%%%%%%%%%%%%%%%%%
let us imagine that two identical 
replicas
(having exactly the same set $J_{ij}$s) 
 of a spin glass sample are allowed to relax thermally below $T_c$, 
starting from two different
random (paramagnetic) initial states. Then these two replicas
(labeled by $\mu$ and $\nu$, say) will
settle in two different valleys, each being characterized by a local
value of the order parameter and the corresponding
 overlap parameters $q_{\mu\nu}$
having a sample-specific distribution
%\begin{equation}
%%%%%%%%%% corr-ref-3-4 %%%%%%%%%%%%%%%%%%
%%%%%%%%%% corr-ref-3-4 %%%%%%%%%%%%%%%%%%%% 
%$\begin{equation}
$$
P_{J}(q) = \sum_{\mu,\nu} e^{-\frac{F_{\mu} + F_{\nu}}{T}}
%P_{J}(q) = \sum_{\alpha,\beta} e^{-\frac{F_{\alpha}}{T}}
%e^{-\frac{F_{\beta}}{T}} 
\delta(q-q_{\mu\nu}); 
$$
\begin{equation}
P(q) = \int \prod_{i>j} dJ_{ij} \rho(J_{ij})P_{J}(q). 
\label{Pq-2}
\end{equation}

\noindent
Here the subscript $J$ denotes a particular
sample with a given realization of quenched random 
interactions ($J_{ij}$s)
between the spins, and finally averaging
$P_{J}(q)$ over disorder distribution $\rho (J)$ in (2) or (3) 
one gets $P(q)$. Physically,  
%Assuming self-averaging behaviour, %%%%%%corr-r1-4 %%%%%%
%Averaging $P_{J}(q)$ over disorders, one gets  %%%%%%corr-r1-4 %%%%%%
%$P_{J}(q)$ may be replaced by its average over disorders in  %%%%%%corr-r1-4 %%%%%%
% thermodynamic limit, i.e., by  %%%%%%corr-r1-4 %%%%%%
%$$P(q) = \int \prod_{i>j} dJ_{ij} \rho(J_{ij})P_{J}(q),$$ 
%\noindent
%where $\rho(J_{ij})$ is the distribution function from which $J_{ij}$s are
%drawn.  
$P(q)$ gives the probability distribution for the two pure states to have 
an overlap $q$, assuming that the probability of reaching any pure
state $\mu$ 
starting from a random (high temperature) state 
is proportional to the
thermodynamic weight $\exp{\{-F_{\mu}\}}$ of the state $\mu$. 
%%%%%%%%%%%%%%%%%%%%%%%%%%%%%%%%%%%%%%%%%%%%%%%%%%%%%%%%%%%%%%%%%%%%%%%% 
The other picture is due to the droplet model 
of short-range spin glass
(Bray and Moore 1984, Fisher and Huse 1986),
where there is no divergence in the typical free energy
barrier height and the relevant timescale is taken to be 
that of crossing the free energy barrier of formation of a
typical droplet of same (all up or all down) spins. 
Based on certain scaling ansatz, this picture leads to 
 a logarithmically 
decaying (with time) self-correlation function for the spins 
below the freezing temperature $T_{c}.$ 

The issue of
the validity of the mean-field picture 
(of the replica symmetry-breaking)
in the context of the 
real-life spin glasses, where the interactions are
essentially short-range, 
is far from settled (see e.g., Moore et al 1998, 
Marinari et al 1998,
Krzakala et al 2001 and Gaviro et al 2006). However, the
effective Hamiltonian (cost function) for 
many other optimization problems
may contain long-range interactions and may
even show the replica symmetry-breaking behavior shown 
in the Graph Partitioning Problem (Fu and Anderson 1986). Of course,
no result of QA for such system (for which replica symmetry-breaking 
is shown explicitly) has been
reported yet. The successes of QA reported so far are mostly
for short-range systems. Thus, the scope of quantum annealing
in those long-range systems still stands out as an interesting
open question.    

%RS is broken (RSB) 
%\subsection{Replica Theory and RSB in the Spin Glass Models} 
%\subsection{Free Energy Minimization} 

%within the ferromagnetic phase (Nishimori 2001). 

\subsubsection{The Traveling Salesman Problem}
\noindent
In the traveling salesman problem 
(TSP), there are $N$ cities placed randomly in a country having a
definite metric to calculate the inter-city distances. A salesman has to
make a tour to cover every city
and finally come back to the starting point. The problem
is to find a tour of minimum length. An instance of the problem is given
by a set $\{d_{ij};$ $i,j = 1,N\}$, 
where $d_{ij}$ indicates the distance between
the $i$-th and the $j$-th city, or equivalently, the cost for going from the
former to the later. We mainly focus on the results of symmetric case, 
where $d_{ij} = d_{ji}$. The problem can be cast into the form where
one minimizes an Ising Hamiltonian under some constraints, as shown below.
A tour can be represented by an
$N\times N$ matrix $\mathbf{T}$ with elements either $0$ or $1$. In a given
tour, if the city $j$ is visited immediately after visiting city $i$, then
$\mathbf{T}_{ij} = 1$, or else $\mathbf{T}_{ij} = 0$. 
Generally an additional constraint
%%%% corr-ref-3-5 %%%%%%%%%%%%%%%%% 
%(which naturally restricts the search 
%to a subspace that contains the
%minimal path, for many metrics)
is imposed that 
one city has to be visited once and only once in a tour. 
Any valid tour with
the above restriction may be
represented by a $\mathbf{T}$ matrix whose each row and each
column has one and only one element equal to $1$ and rest all
are $0$s. For
symmetric a metric, a tour and its reverse tour have the
same length, and it is more convenient to work with an undirected tour matrix
$\mathbf{U} = \frac{1}{2} (\mathbf{T} + \tilde \mathbf{T})$, where 
$\tilde \mathbf{T}$, the transpose of $\mathbf{T}$, represents the reverse of
the tour given by $\mathbf{T}$. Clearly, $\mathbf{U}$ must be a 
symmetric matrix having two
and only two distinct
entries equal to $1$ in every row and every column, no two rows 
being identical, and so is not any two columns. In terms of 
$\mathbf{U}_{ij}$s, the length of a tour can be represented by
%%%% corr-ref-3-3 %%%%%%%%%%%%%%
\begin{equation}
\mathcal{H} = \frac{1}{2}\sum_{i,j=1}^{N} d_{ij}\mathbf{U}_{ij}. 
\end{equation} 
\noindent
One can rewrite the above Hamiltonian in terms of Ising spins 
$S_{ij}$s as
\begin{equation}
\mathcal{H}_{TSP} = \frac{1}{2}\sum_{i,j=1}^{N} d_{ij}\frac{(1 + S_{ij})}{2}
\label{TSP-Ham}
\end{equation}
\noindent
where $S_{ij} = 2\mathbf{U}_{ij} - 1$ are the Ising spins. 
The Hamiltonian is similar
to that of a non-interacting Ising spins on an $N\times N$ lattice, with
random fields $d_{ij}$ on the lattice points $\{i,j\}$. The frustration
is introduced by the global constraints on the spin configurations
in order to conform with the structure of the matrix $\mathbf{U}$ discussed
above. The problem is to find the ground state of the Hamiltonian subject
to these constraints. 
There are $N^2$ Ising spins,
which can assume $2^{N^2}$ configurations in absence of any constraint,
but the constraint here reduces the number of valid configurations to that
of the number of distinct tours, which is $(N!)/2N$. 

%The hardness of the problem depends on the nature of the metric $d_{ij}$,
%or rather, on the correlation among the variables $d_{ij}$s.
Mainly two distinct classes of TSP 
are studied: one with an Euclidean $d_{ij}$ 
in finite dimension (where $d_{ij}$ are strongly correlated through
triangle inequalities, which means, for any three cities $A$, $B$ and $C$, the sum of any two
of the side $AB$, $BC$ and $CA$ must be greater than the remaining one), 
and the other with random $d_{ij}$ in infinite dimension.

In the first case, $N$ cities are uniformly distributed 
within a hypercube in a $d$-dimensional Euclidean space. 
Finding a good approximation for large $N$ is 
easier for this case, since the problem is finite-ranged. Here a 
$d$-dimensional neighborhood is defined for each city, and the problem
can be solved by dividing the whole hypercube into a 
number of smaller pieces and
then searching for the least path within each smaller 
part and joining them back. The
correction to the true least path will be due to the unoptimized
connections across the boundaries of the subdivisions. 
For a suitably made division (not too small), this
correction will be of the order of the surface-to-volume ratio
of each division, and thus will tend to zero in the $N \rightarrow \infty$ 
limit. This method, known as ``divide and conquer'', forms a
reasonable strategy for solving approximately 
such finite-range optimization problems
(including finite-range spin glasses) in general. 
In the second case, $d_{ij}$s are assigned completely randomly, with no
geometric
(e.g., Euclidean) 
correlation between them. The problem in this case becomes
more like a long-range spin glass. A self avoiding walk representation
of the problem was made using an $m$-component vector field, and the
replica analysis was done (Mezard et al 1987) for finite temperature, 
assuming replica symmetry ansatz to hold. Moreover, true breaking of 
ergodicity may occur only in infinite systems, not in any finite
instance of the problem.
The results, when extrapolated to zero-temperature,
do not disagree much with the numerical results (Mezard et al 1987). 
The
stability of a replica symmetric solution has not yet been proved for low
temperature region. However, numerical results of thermal annealing for 
instances of size $N = 60$ to $N = 160$ yielded many near optimal tours, and
the corresponding overlap analysis shows a sharply peaked distribution, whose
width decreases steadily with increase in $N$. This indicates the
existence of a replica symmetric phase for the system (Mezard et al 1987).  

An analytical bound on the average
(Normalized by $N^{1/2}$) value of optimal path-length per city 
($\Omega$) calculated
for TSP on 2-dimensional Euclidean plane  
%(where one
% calculates the average distance of successive cities within the 
%strips of a given width, and then optimize that with respect to the
%width), 
has been found to be $ 5/8 < \Omega < 0.92$ (Bearwood et al 1959). Careful
scaling analysis of the numerical results obtained so far indicates the
lower bound to be close to $0.72$ 
(Percus and Martin 1996, Chakraborti and Chakrabarti 2000).  

Simulated (thermal) annealing of a Euclidean TSP on a square 
having length $N^{1/2}$ 
(so as to render the average nearest neighbor distance independent of $N$)
has been reported (Kirkpatrick et al 1983). 
In this choice of length unit, the optimal tour length per
step ($\Omega$) becomes independent of $N$ for large $N$. 
 Thermal annealing rendered $\Omega \le 0.95$ 
for $N$ up to $6000$ cities. 
This is much better than what is obtained by the so 
called ``greedy heuristics'',
(where being at some city in a step, one moves to the nearest city not in
the tour in the next step) for which $\Omega \sim 1.12$ on average.
Later we will see that (Sec. IIIA) that quantum annealing
can do even better than thermal annealing in context of random TSP.

%%\section{Optimizations, Thermal Annealing and Spin Glasses}

%\subsection{Simulated Thermal Annealing}

To summarize, when cast in some energy minimization problem, the
combinatorial optimization problems may exhibit glassy behavior during
thermal annealing. Even replica symmetry-breaking behavior may be observed 
(like in case of Graph Partitioning Problem, see Fu and Anderson 1986), 
since the underlying Hamiltonian need not be short-range, 
and the constraints can
bring frustrations into the problem. One can intuitively conclude that
thermal annealing or other heuristics would not be able to solve such 
problems easily to a good approximation within reasonable time. 
%However, the task of determining the ground state of a finite-ranged
%spin glasses can be quite daunting and one such problem
%has explicitly been shown to belong
%to the NP-complete class (Barahona 1982).
Moreover, practically nothing 
can be said about the time
required to solve the worst case instance exactly. 
Specifically, in some cases, where
good solutions are thermodynamically very 
insignificant in number and there is
no monotonic gradient towards them, the entropy might make a
classical search exponentially
difficult, though the landscape might still remain completely ergodic.
Later we will see quantum searches can bring about spectacular
improvements in some such cases (see Sec. IIIB and Fig. \ref{qan}).

%%%%%%%%%%%%%%%%%%%%%%%%%%%%%%%%%%%%%%%%%%%%%%%
\subsection{Quantum Spin Glasses and Annealing}
%%%%%%%%%%%%%%%%%%%%%%%%%%%%%%%%%%%%%%%%%%%%%%%
\noindent
In QA one adds a kinetic (tunneling) term to the
interaction part of the classical glass Hamiltonian. The
object that results in, is called a quantum spin glass.
The knowledge of the phase-diagram of a quantum spin glass
is crucially important for its annealing, as it gives
an idea of the location of the quantum critical points on the
phase diagram, and thus offers a guideline to choose the proper
kinetic terms (that maintain a sizable gap)
and the suitable annealing paths 
(see Sec. IIIA and IIIB).

In quantum spin glasses (Chakrabarti 1981, Ishii and Yamamoto 1985,
Bhatt 1998, Rieger 2005, Sachdev 1999, Ye 1993), the order-disorder
transition (i.e., from the frozen phase to the high kinetic energy phase, 
termed
the para phase) 
can be driven both by thermal 
fluctuations as well as by quantum
fluctuations. Quantum spin glasses can be of two types: 
vector spin glasses,
where the
 quantum fluctuations cannot be %%%%%%%%% corr-ref-3-8 %%%%%%%%%
adjusted by changing some laboratory field, and the
other, a classical spin glass perturbed by some quantum
tunneling term, where the quantum fluctuations are controlled
through, say, a transverse laboratory field.

The amount of the quantum fluctuations being adjustable, this transverse Ising
spin glass (TISG) model is perhaps the simplest model in which the quantum
effects in a random system can be and have been studied extensively and
systematically (Chakrabarti et al 1996). 
Here we will focus only on TISG, since
the reduction
 of the quantum fluctuations is the key feature required for
quantum annealing. 

The 
%recent spectacular upsurge 
 interest %%%%%% corr-r1-6 %%%%%%%
in the zero-temperature quantum spin glass phases in the
TISG models have been have been complimented all along by the experimental
studies in several systems which have been established to be represented
accurately by transverse field Ising model (TIM). Recent discovery of the
compound material LiHo$_{x}$Y$_{1-x}$F$_4$ with the magnetic Ho ion 
concentration $x = 0.167$ (Aeppli and Rosenbaum 2005, Wu et al 1991, 
Wu et al 1993a; see also Brooke et al 2001, Silevitch et al 2007) 
representing accurately a random long-range transverse Ising system,
has led to renewed interest. Here, the strong spin-orbit coupling between 
the spins and the host crystals restricts the effective ``Ising'' spins
to align either parallel or anti-parallel to the specific crystal axis.
An applied magnetic field, transverse to the preferred axis, flips the
``Ising'' spins. This feature, together with the randomness in the spin-spin
interaction, makes it a unique TISG-like system. 
Most interestingly, it has been shown 
that in spite of the presence of
all the three ingredients - 
frustrations, randomness and the long-range (dipolar) interactions, 
that are necessary for the formation of a spin glass, the 
spin glass phase of LiHo$_{x}$Y$_{1-x}$F$_4$ is actually destroyed
by any finite transverse field (Schechter and Laflorencie 2006). 
This indicates 
the effectiveness of quantum tunneling in
the exploration of a rugged PEL with formidable potential energy barriers.
%\subsection{Model}
%\noindent
%%%%%%%%%%% corr-ref-3-9 %%%%%%%%%%%%
 The TISG model described here, is given
by the Hamiltonian
\begin{equation}
{\mathcal H} = -\sum_{i>j}^{N} J_{ij}S_{i}^{z}S_{j}^{z} - \Gamma\sum_{i}^{N}
S_{i}^{x},
 \label{SK-Q}
\end{equation}
\noindent where $\Gamma$ denotes the tunneling strength at each site
and $J_{ij}$s are distributed randomly following the distribution 
$\rho(J_{ij})$ given by (\ref{Gaussian}) or (\ref{pls-mns-J}).
Generally, we denote the strength of the 
quantum kinetic term 
by $\Gamma.$

The unique interest in such quantum spin glass system comes from the
possibility of much faster crossing of the  high barriers occurring 
 in the potential energy landscapes of the classical spin glasses  
by means of quantum tunneling induced by the transverse field,
compared to that done thermally by
scaling such barriers with the temperature.
%In classical
%systems, the overriding of an energy barrier becomes exponentially harder
%for thermal fluctuations as the height grows. However,
%for the narrow barriers, quantum fluctuations help tunneling through.
%can make a system tunnel through such a barrier, if its width is
%infinitesimally small. 

%\subsection{Phase Diagram}
%\noindent
The phase transitions in quantum spin glasses can 
be driven both by thermal and quantum fluctuations as mentioned before
and the equilibrium phase diagrams
also indicate how the optimized solution (in SG phase) can be obtained
either by tuning of the temperature $T$ or the tunneling field $\Gamma,$
or by both. We will show later (in context of quantum annealing) that
reaching the phase by tuning $\Gamma$ may often be more advantageous 
(faster) than that by tuning $T$. 

The short-range version of this TISG model was first studied by
Chakrabarti (1981), and the long-range version, discussed here,
was first studied by Ishii and Yamamoto (1985).
Several analytical studies have been made to obtain the phase
diagram of the transverse Ising SK model (Miller and Huse 1993).
%giving in particular the zero-temperature critical field).  
%%%%%%%% corr-r1-8 %%%%%% 
The problem of an SK glass in a transverse
field becomes a nontrivial one due to the presence of non-commuting spin
operators in the Hamiltonian. This leads to a dynamical frequency-dependent
 self-interaction for the spins.

%\subsubsection{Mean Field Estimate}

\begin{figure}[h]
\resizebox{6.0cm}{!}{\rotatebox{0}{\includegraphics{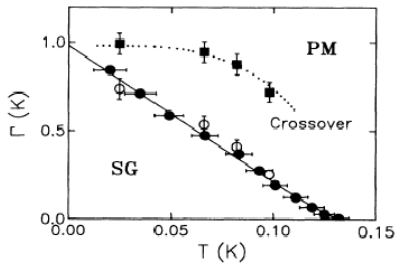}}}
\caption{{\footnotesize{
Phase diagram of LiHo$_{0.167}$Y$_{0.833}$F$_4$
according to the dynamical measurements
(filled circles) and 
 of the nonlinear susceptibility measurements
(open circles). Filled squares indicate the freezing boundary
obtained from AC susceptibility measurements 
(taken from Wu et al 1993b).}}} 
\label{Exp-Phsd}
\end{figure}

One can study an effective-spin Hamiltonian for the SK model
in a transverse field within the mean field framework very easily. 
The spin glass order parameter in a classical SK model 
is given by a random `mean field'
$h(r)$ having Gaussian distribution (see Binder and Young 1986)
\begin{equation}
q = \int_{-\infty}^{+\infty} dr e^{-r^{2}/2} \tanh^{2}{(h^{z}(r)/T)}; 
\quad h^{z}(r) = J\sqrt{q}r + h^{z}
\label{mf-q}
\end{equation}
\noindent
where $h^{z}$ denotes the external field (in $z$ direction), 
the mean-field $h(r)$ being also in the same direction. In the 
presence of the transverse field, as in (\ref{SK-Q}), $h(r)$ has 
components both in $z$ and $x$ directions
%%%% corr-ref-3-3 %%%%%%%%%%%%%%
\begin{equation}
\vec{h}(r) = -h^{z}(r)\hat{z} -\Gamma\hat{x}; 
\quad h(r) = \sqrt{h^{z}(r)^{2} + \Gamma^{2}},
\end{equation}
\noindent
and one replaces the ordering term $\tanh^{2}{(h(r)/T)}$ in
(\ref{mf-q}) by its component
$[|h^{z}(r)|/|h(r)|]^{2}\tanh^{2}{(|h(r)|/T)}$ in the $z$-direction. Putting
 $h^{z} = 0$ and $q\rightarrow 0$ one gets the phase boundary equation as
(see Chakrabarti et al 1996)
\begin{equation}
\frac{\Gamma}{J} = \tanh{\left(\frac{\Gamma}{T}  \right)}.
\label{E-30} 
\end{equation}
\noindent
This gives $\Gamma_{c}\quad (T=0) = J = T_{c}\quad (\Gamma=0)$ and a
phase diagram qualitatively similar to the experimental one shown in
Fig. \ref{Exp-Phsd}. 

%\subsubsection{Monte Carlo Studies}
%\noindent 
Several Monte Carlo studies have
been performed 
for the SK spin
glass model in transverse field applying \index{Suzuki-Trotter!formalism}
the Suzuki-Trotter formalism (see Appendix 1), mapping a
$d$-dimensional quantum Hamiltonian 
 to an effective $d+1$ dimensional anisotropic classical Hamiltonian
(see also Hatano and Suzuki 2005). The 
 partition function
gives the effective
classical Hamiltonian in the $M$th Trotter approximation as

%%%% corr-ref-3-3 %%%%%%%%%%%%%%
\begin{equation}
{\mathcal H} = \sum_{i>j}^{N}\sum_{k}^{M} K_{ij}S_{ik}S_{jk}
 -\sum_{i}^{N} \sum_{k}^{M} K S_{ik}S_{ik+1},
\end{equation} 
\noindent with
\begin{equation}
 K_{ij} = \frac{J_{ij}}{MT} ;\qquad K = \frac{1}{2}
\ln{\coth{\left( \frac{\Gamma}{MT} \right),}} 
\label{E-32}
\end{equation}
\noindent where $S_{ik}$ denotes the Ising spin defined on the lattice site
($i$, $k$), $i$ denoting the position in the in the original SK model and
$k$ denoting the position in the additional Trotter dimension.
Although the equivalence between classical and the quantum model
holds exactly in the limit $ M\rightarrow \infty , $ one can always make an
optimum choice for $M$. The equivalent classical Hamiltonian has been
studied using standard Monte Carlo technique. The numerical estimates of the
phase diagram etc. are reviewed in details in (Bhatt 1998, Rieger 2005).
Ray et al (1989) 
took $\Gamma << J$ and their results indeed indicated
a sharp lowering of $T_{c}(\Gamma)$. Such sharp fall of $T_{c}(\Gamma)$ with
large $\Gamma$ is obtained in almost all the 
theoretical studies of the phase
diagram of the model (Miller and Huse 1993; Ye et al 1993; 
see also Bhatt 1998 and Rieger 2005).
Quantum Monte Carlo (Alverez and Ritort 1996)
as well as real-time Schr\"{o}dinger evolution 
(the true dynamics given by the time-dependent
Schr\"{o}dinger equation)
studies 
of SK spin glass in transverse field 
were made (Lancaster and Ritort, 1997).   

In the Hamiltonian for the EA spin glass in presence of a
transverse field, given by (\ref{SK-Q}), the random interactions
are restricted among the nearest neighbors and satisfy a Gaussian
distribution with zero mean and variance $J$, as given by Eq. (\ref{Gaussian}).
Here, the variation of correlations in the equivalent $(d+1)$ dimensional
classical model fitted very well (Guo et al 1994) with the scaling fit
with a unique order parameter and a
critical interval corresponding to a phase diagram whose features are
similar to those discussed above (see also 
Chakrabarti et al 1996, Bhatt 1998 and Rieger 2005). 

%\subsubsection{Experimental Studies}
%\noindent
As discussed earlier in this section, LiHo$_{x}$Y$_{1-x}$F$_{4}$
with $x = 0.167$ provides a spin glass system, for which the external
magnetic field transverse to the preferred axis scales like the square root
of the tunneling field $\Gamma$ in (\ref{SK-Q}). With increasing
transverse field, the glass transition temperature decrease monotonically,
as shown in Fig. \ref{Exp-Phsd}.  
%%%\subsubsection{Ergodicity in Quantum Spin Glasses}
%\subsubsection{Ground State Structure}
%\noindent 
%The question of existence of replica-symmetric ground state in
%quantum spin glasses has been studied extensively in recent years. Replica
% symmetry restoration (even partial)
%is a quantum phenomenon arising due to the quantum
%tunneling between the classically `trapped' states separated by infinitely
%high (but infinitesimally narrow) barriers in the free energy surface.
A quantum tunneling term allows for overlap between two classically
localized states and the dynamics near the ground states of such glasses
show better ergodicity properties.  
In order to
investigate this aspect of quantum spin glasses, 
one can study the overlap distribution function $P(q)$ given by Eq. 
(\ref{Pq-2}). 

%(\ref{Pq-1}). 
%In quantum glass problem one can study this 
%overlap distribution $P_{N}(q);$
If the ergodicity is recovered, 
at least for a part of the
phase diagram,
%replica symmetric ground states exists, 
the above function should tend
to a delta function form, 
peaking at some finite value of the order-parameter $q$
in thermodynamic limit. 
In the para-phase, of course, the
 distribution becomes a delta function 
 at $q = 0$ for the infinite system. In spite of several
investigations 
(
see e.g., Ray et al 1989, Thirumalai and Kirkpatrick 1989, 
Goldschmidt and Lai 1990, Chakrabarti et al 1996, 
Kim and Kim 2002) the point 
of replica symmetry restoration in spin glasses
by quantum fluctuation
is not settled yet. However, slow withdrawal
(see Eq. \ref{adia-cond} for the characteristic slowness) of 
the tunneling field in these quantum spin glasses can help 
annealing the system close to 
the ground state of the classical spin glass eventually, as 
 has been described in the next section.
 
\section{Quantum Annealing\label{QOPT}}
\noindent
In the previous sections we have seen how thermal fluctuations can be utilized
to devise fast heuristics to find an approximate ground state of a glassy
system, or equivalently, a near-optimal solution to combinatorial 
problem, whose 
cost-configuration landscape has glassy behavior due to the 
occurrence of many local minima. There are two aspects of an optimization
problem which might render thermal annealing to be a very
ineffective one. 
First, in a glassy landscape, there
may exist very high cost/energy barriers around local minima which does not 
correspond to a reasonably low cost (see Fig. \ref{QA-SA}). 
In the case of infinite-range problems, these
barriers might be proportional to the system-size $N$, and thus diverge in
thermodynamic limit. Thus there might occur many unsatisfactory local minima,
any of which can trap the system for very long time 
(which actually diverges in thermodynamic limit for infinite-range systems) 
in course of annealing. The second problem is the entropy itself. 
The number of configurations grow very fast 
with the number of variables
(roughly exponentially; $n$ Ising spin can be 
in $2^{n}$ configurations) and a classical system can only assume
one configuration at a time and 
%(unlike a quantum system, which 
%can be in a superposition of all of them at a given instant). 
unless there is a gradient that broadly guides the system
towards the global minimum form any point in the configuration space,
the search has to involve visiting any substantial fraction
of the configurations. Thus a PEL without 
a guiding gradient poses a 
problem which is 
clearly of an exponential or higher order
in complexity
 (depending how the size of the
configuration space scales with the system-size $N$), and 
the CA algorithms can do
no
better than a random search algorithm. This is the case for
golf-course type potential-energy landscape (where there is a sharp
potential minimum on completely flat PEL) 
(see Sec. IIIB, Fig. \ref{qan}). 
One can imagine that 
quantum mechanics might have some solutions to both 
these problems, at least, to some extent. 
This is because quantum mechanics can introduce
classically unlikely tunneling paths even through very high barriers if they
are narrow enough (Ray et al 1989; see also Apolloni et al 1989, 1990). 
This can solve the ergodicity problem to some extent, as
discussed earlier. Even in places where ergodicity breaking does not take 
place in true sense, once the energy landscape contains high enough barriers
(specially for infinite-ranged quenched interactions), quantum tunneling
may provide much faster relaxation to the ground state 
(Santoro et al 2002, Marto\u{n}\'{a}k et al 2002; see also  
Santoro and Tosatti 2007, Somma et al 2007). 
In addition, a quantum mechanical wave function can delocalize
over the whole configuration space (i.e., over the potential
energy landscape) if the kinetic energy term is high enough. 
Thus it can actually
``see" the whole landscape simultaneously at some stage of annealing. 
These two
aspects can be expected to improve the search process when employed properly. 
In fact such improvements can indeed be achieved in certain situations, though 
quantum mechanics is not a panacea for all such diseases
as ergodicity breaking, spin glass behavior etc, and certainly
has its own inherent limitations. What intrigues one is
the fact that the limitations due to quantum nature of an algorithm
are inherently different from those faced by its classical counterpart,
 and thus it is not yet clear in general which wins when. Here 
we discuss some results regarding the quantum heuristics and
some of their general aspects understood so far. For more 
detailed reviews of the subject we refer to
 the articles in Das and Chakrabarti (2005) and Santoro and 
Tosatti (2006).

Some basic aspects of QA can be understood from the simple case of 
QA in the context of a double-well potential 
(Stella 2005, Battaglia et al 2005). Typically a particle 
in a double-well consisting of a shallower but wider well and a deeper 
but narrower well is annealed (it is likely that the deeper well, i.e., the
target state, is narrower, otherwise the searching becomes easier 
even classically). 
The kinetic energy (inverse mass) is
tuned from a very high value to zero linearly within a time $\tau$. For
a very high value of initial
kinetic energy, the wave function, 
which is the ground state, is delocalized more or less over the
whole double-well. As kinetic
energy is reduced but still quite high, the ground state corresponds to
a more pronounced peak on the shallower minimum, since it is wider. This
is because at this stage, to obtain the minimal (ground state) energy,
it is more effective to minimize the 
kinetic energy by localization over a wider
space, rather than minimizing the potential energy by localizing in the 
deeper well. However, as kinetic energy is tuned down further, the potential
energy term becomes dominating, and the ground state has a taller peak 
around the deeper minimum. The evolving wave function can roughly
follow this ground state structure all the way and finally settle to the
deeper minimum if the annealing time $\tau$ is greater than some $\tau_c$.
When $T < T_{c}$ 
it fails to tunnel from its early state localized in the shallower well
to the deeper well as the kinetic energy is decreased. This result is 
qualitatively the same for both real-time and quantum Monte Carlo annealing, 
excepting for the fact that the $\tau_c$'s are different in the two cases.     

The realization of QA consists of employing
adjustable quantum fluctuations into the problem
instead of a thermal one (Kadowaki and Nishimori 1998,
Amara et al 1993, Finnila et al 1994). 
In order to do that, one needs to introduce an
artificial quantum kinetic term $\Gamma(t) \mathcal{H}_{kin}$, 
which does not commute
with the classical Hamiltonian $\mathcal{H}_C$ representing the cost function.
The coefficient $\Gamma$ is the parameter which controls the quantum
fluctuations.
The total Hamiltonian is thus given by 
\begin{equation}
\mathcal{H}_{tot} = \mathcal{H}_C + \Gamma(t)  \mathcal{H}_{kin}.
\label{Total-Ham}
\end{equation} 
\noindent  
The ground state of $\mathcal{H}_{tot}$ is a superposition of the 
eigenstates of $H_C$. 
%which basically represents the classical configurations.  
For a classical Ising Hamiltonian of the form (\ref{spg-ham}), the 
corresponding total quantum Hamiltonian might have the form (\ref{SK-Q}),
where $\mathcal{H}_C = -\sum_{i>j}i^{N} S_{i}^{z}S_{j}^{z}$ 
%$S_{i}^{z}$ being the$z$-component of 
%Pauli's spin representing the $i$-th Ising spin, 
and
$\mathcal{H}_{kin} = -\sum_{i}^{N}S_{i}^{x}$. 
%where $S_{i}^x$ is the 
%$x$-component of Pauli's spin, representing the on-site kinetic
%(spin flip) term.
Initially $\Gamma$ is kept very high so that the $\mathcal{H}_{kin}$
dominates and the ground state is trivially a uniform superposition of
all the classical configurations. One starts with that uniform superposition
as the initial state, and slowly decreases $\Gamma$ following some annealing
schedule, eventually to zero. 
If the process of decreasing is slow enough, the 
adiabatic theorem of
quantum mechanics (Sarandy et al 2004) 
assures that the system will always remain at the 
instantaneous ground state of the evolving Hamiltonian $\mathcal{H}_{tot}$.\
When $\Gamma$ is finally brought to zero, 
$\mathcal{H}_{tot}$ will coincide with the original classical Hamiltonian
$\mathcal{H}_{C}$ and the system will be found in the ground state of it,
as desired. 
The special class of QA algorithms, where 
strictly quasi-stationary
or adiabatic evolutions are employed are also known as Quantum 
Adiabatic Evolution algorithms (Farhi et al 2001).

Two important questions are how to choose an appropriate $\mathcal{H}_{kin}$
and how slow the evolution needs to be in order to 
assure adiabaticity. According to the adiabatic theorem of
quantum mechanics, for a
non-degenerate spectrum with a gap between the ground state and
first excited state, the adiabatic evolution is assured if the evolution time
$\tau$ satisfies the following condition-
\begin{equation}
\tau >> \frac{ 
|\langle \dot{\mathcal{H}_{tot}} 
\rangle|_{max}}{\Delta^2_{min}},
\label{adia-cond}
\end{equation}  
\noindent
where
%\begin{eqnarray}
$$
|\langle\dot{\mathcal{H}_{tot}}\rangle|_{max} =
\max_{0 \le t\le \tau}\left[ 
\left|\left\langle 
\phi_{0}(t)\left|\frac{d\mathcal{H}_{tot}}{ds}\right|\phi_{1}(t) 
\right\rangle\right|\right], $$
%\nonumber \\
\begin{equation}
\Delta^2_{min} = \min_{0 \le t \le \tau}\left[ \Delta^2(t)\right];   
\quad s = t/\tau;\quad 0\le s \le 1,
\label{adia-factors}
\end{equation}
\noindent
$|\phi_{0}(t)\rangle$ and $|\phi_{1}(t)\rangle$ being respectively 
the instantaneous ground state and the first excited state of the total
Hamiltonian $\mathcal{H}_{tot}$, and $\Delta(t)$ the instantaneous gap
between the ground state and the first excited state energies 
(see Sarandy et al 2004). One may wonder that while entering the
ordered phase ($\Gamma < \Gamma_{c}$) from the para phase 
($\Gamma > \Gamma_c$) in course of annealing, 
the gap $\Delta$ may vanish 
at the phase boundary ($\Gamma = \Gamma_{c}$) in the 
$N\rightarrow\infty$ limit. 
In fact, in such a case, 
QA cannot help anyway in finding the ground state 
of an infinite system.
However, for any finite sample, this gap is very
unlikely to vanish for a random system, and QA
may still work out nicely.

However, it is impossible to follow, 
even for a finite $N$,
the evolution of
a full wave function in a classical computer using polynomial 
resources in general, since it requires tracking   
the amplitudes of all the basis vectors 
(all possible classical configurations), whose number
grows exponentially with system-size $N$. Such an adiabatic evolution
may be realized within polynomial resources only if one can employ a 
quantum mechanical system itself to mimic the dynamics. 
However, one may employ quantum Monte Carlo methods to simulate some
dynamics (not the real-time quantum dynamics) 
to sample the ground state (or a mixed state at low enough temperature) 
for a given set of parameter values of
the Hamiltonian. Annealing is done by reducing the strength $\Gamma$ of the
quantum kinetic term in the Hamiltonian from a very high value to zero
following some annealing schedule in course of simulation. In case of such
Monte Carlo annealing algorithm, there is no general bound on 
success time $\tau$ like
the one provided by the adiabatic theorem for true Schr\"{o}dinger evolution
annealing. Here we will separately discuss the results of real-time QA and
Monte Carlo QA. Apart from these quasi-stationary quantum annealing
strategies, where the system always stays close to some stationary state
(or low-temperature equilibrium state), there may be cases where 
quantum scatterings (with tunable amplitudes) 
are employed to anneal the system (Das et al 2005). 
\subsection{Quantum Monte Carlo Annealing}
\noindent
In quantum Monte Carlo annealing, one may employ 
either a finite (but low) temperature
algorithm, or a zero-temperature algorithm. Most of
the Monte Carlo QA (Das and Chakrabarti 2005, 
Santoro and Tosatti 2006) %%%% corr-r2-1 %%%%
are done using a finite temperature Monte Carlo, 
namely, the Path Integral Monte Carlo 
(PIMC), since its implementation is somewhat simpler %%%%% corr-r2-1 %%%%%
than that of other zero-temperature Monte Carlo methods . %%%%%% corr-r2-1 %%%%

Among the other zero-temperature Monte Carlo
methods used for annealing are the
zero-temperature transfer-matrix Monte Carlo 
(The chapter by Das and Chakrabarti in Das and Chakrabarti, 2005) 
and the Green's function Monte Carlo (Santoro and Tosatti 2006). 
However, these algorithms suffer severely from different drawbacks, which
renders them much slower than PIMC algorithms in practice.

The Green's function Monte Carlo effectively simulates the real-time evolution
of the wave function during annealing. But to perform sensibly, it often 
requires a guidance that depends on a priori knowledge of the wave function. 
Without this guidance it may fail miserably (Santoro and Tosatti 2006). 
But such a priori knowledge is very unlikely to be available in the case of
random optimization problems, and hence so far the scope  
for this algorithm seems to be very restricted.

The zero-temperature transfer-matrix Monte Carlo method, on the other hand, 
samples the ground state of the instantaneous Hamiltonian 
(specified by the given value of the parameters at that instant) using a 
projective method, where the Hamiltonian matrix
itself (a suitable linear transformation
of the Hamiltonian that converts into a positive matrix, 
in practice)  
is viewed as the transfer-matrix of a finite-temperature classical system of
one higher dimension (Das and Chakrabarti 2005). But the sparsity of
the Hamiltonian matrix for systems with local kinetic terms 
leaves the classical
system highly constrained and thus very difficult to simulate efficiently
for large system sizes.

The PIMC has so far been mostly used for QA.
The basic idea of path-integral Monte Carlo rests on Suzuki-Trotter
formalism (see Appendix 1), 
which maps the partition function of a $d$-dimensional 
quantum Hamiltonian $\mathcal{H}$ to that of an 
effective classical Hamiltonian 
$\mathcal{H}_{eff}$ in ($d+1$)-dimension. 
Quantum annealing of a Hamiltonian $\mathcal{H}_{tot}$ using
PIMC consists of mapping $\mathcal{H}_{tot}$ to its equivalent classical one
and simulate it at some fixed low temperature so that thermal fluctuations are
low. The quantum fluctuations are reduced from some very high initial value to
zero finally through reduction of $\Gamma$  in course of simulation. Clearly,
the simulation dynamics is not the true Schr\"{o}dinger evolution of the 
system, and also it
cannot simultaneously see the whole configuration space as
does a delocalized wave function. It feels the landscape locally and makes
moves just like a classical system. The first attempt of quantum annealing
using PIMC was done by 
Kadowaki and Nishimori (1998) for solving TSP 
and an extensive use of the technique 
to explore multitude of problems has been made by the group of 
Santoro and Tosatti (2006). Here we will discuss some
PIMC quantum
annealing results briefly for few different systems.   

%%%%%%%%%%%%%%%%%%%%%%%%%%%%%%%%%%%
%\vspace{0.4cm} 

%\noindent
%(i) {\textit{2D EA Spin Glass}}

%\vspace{0.45cm}
%%%%%%%%%%%%%%%%%%%%%%%%%%%%%%%%%%%%
%%%\subsubsection{2D EA Spin Glass}
\subsubsection{A Short-Range Spin Glass}
\noindent
Quantum annealing of an EA spin glass in 2-dimension (square lattice)
using transverse field (see Eq. (\ref{SK-Q})) for
large lattice size
(up to $80\times 80$) using PIMC turns out to be much more efficient
compared to thermal annealing (CA) 
of the same system in finding the approximate
ground state (Santoro et al 2002, Marto\u{n}\'{a}k et al 2002). 
The quantity which is measured is the residual
energy $\epsilon_{res}(\tau) = E(\tau) - E_{0}$, $E_{0}$ being the true
ground state energy of the 
finite system (calculated by the Spin Glass Server
using the so called Branch-and-Cut algorithms;
 see $http://www.informatik.uni-koeln.de/ls_juenger/research/spinglass/$) 
and $E(\tau)$ being the final energy of the
system after reducing the transverse field strength $\Gamma$ linearly 
within time $\tau$ from some suitably large initial value to zero. Here $\tau$
is a fictitious time given by the number of Monte Carlo steps. 

Classically,
for a large class of frustrated disordered system it can be shown,
using very general arguments that the residual energy decreases following
some power law in the logarithm of the annealing time $\tau$, namely,
$\epsilon_{res} \sim (\log{\tau})^{-\zeta}$, with $\zeta \le 2$
(Huse and Fisher 1986). In PIMC, the partition function of a $d$-dimensional
quantum system is mapped to an equivalent $d+1$-dimensional classical system
(known as Suzuki-Trotter mapping). The effective $d+1$-dimensional system,
is obtained by replicating the classical part of the
original system (say, the Ising interaction part of the
transverse Ising system without the transverse field term) 
with all its interactions
(including the disorders) intact
along the extra higher dimension. The coupling between the spin
in different replica depends on the quantum kinetic term in the original
 $d$-dimensional system. 

%effectively deals with a 
%classical frustrated system
%but with a rather uncommon anisotropy with perfectly correlated
%randomness along the extra (Trotter) dimension. 
However, the PIMC annealing results
show that the quantum effect (taken into account through Suzuki-Trotter
mapping) does not change the basic 
relaxation behavior of $\epsilon_{res}(\tau)$. But still a dramatic
improvement in evaluation time is achieved since it turns
out that the value of the exponent $\zeta$ can be  much higher ($\zeta = 6$)
for QA
than the Huse-Fisher bound of $\zeta \le 2$ for classical annealing.
This is a tremendous improvement in computational 
time if one thinks
in terms of the changes in $\tau$ required to
improve $\epsilon_{res}$ equally
by some appreciable factor in the
respective cases of classical and quantum annealing. An interesting
 asymptotic comparison for results of QA and CA for an $80\times80$ 
lattice shows
that to reach a certain value of $\epsilon_{res}$, the PIMC-QA would take one
day of CPU time (for the computer used) whereas CA would take
about 30 yrs (Santoro et al 2002, Marto\u{n}\'{a}k et al 2002).
The result would not be much different in this case if the real-time
Schr\"{o}dinger evolution would have been followed, as has been
argued using the Landau-Zener cascade tunneling picture (Santoro et al 2002,
Marto\u{n}\'{a}k et al 2002). 

Landau-Zener tunneling theory gives an estimate of the 
probability of tunneling non-adiabatically from a lower
to a higher level when the system
encounters an avoided level crossing between the levels during time evolution. 
Let the gap between two energy 
levels $|a(t)\rangle$ and $|b(t)\rangle$ of a
time-dependent Hamiltonian vary linearly with time (gap $\Delta = \alpha t $)
and encounter an avoided level crossing. Here the levels are
energy levels of the classical part of the Hamiltonian
(say, the potential energy levels of an Ising model).  
Let the system be made to
evolve in such a way that the characteristic
time it spends while passing through the crossing region is $\tau$. Let there
be a quantum 
tunneling field $\Gamma$ that induces transitions between the levels. 
Then if the system evolved is at the lower branch
$|a\rangle$ before encountering the avoided level crossing, the 
probability that it tunnels to the higher branch $|b\rangle$ while passing
through the crossing decreases with the time $\tau$ as 
$P(\tau) = \exp{(\tau/\tau_{\Gamma})}$, where 
$\tau_{\Gamma} = (\hbar\alpha\Gamma)/(2\pi\Delta_{min}^{2})$, $\Delta_{min}$ 
being the minimum value of the gap attained at the avoided level crossing.  
For spin glass-like systems with non-zero gap, treating multiple
level-crossings, each with small $\Delta$, 
as a cascade of independent Landau-Zener tunneling,
one may argue that residual energy goes as 
$\epsilon_{res} \sim (\log{\tau})^{-\zeta_{Q}}$, where $\zeta_{Q}$ is 
essentially greater than the bound $\zeta\le 2$ for thermal annealing,
and might be as high as 6 (Santoro et al 2002, Marto\u{n}\'{a}k et al 2002). 
%A gap typically goes as $\Delta_{min} \sim \exp{(-d_{ab}/\lambda(\Gamma))}$, 
%where
%$d_{ab}$ is some measure of distance between the 
%two state at the avoided level crossing,
%say, the number of spin-flips to reach from $|a\rangle$ to $|b\rangle.$
%The factor $\labda(\Gamma)$ is the localization length of the wave functuion,
%and typically goes as $\lambda(\Gamma) \sim \Gamma^{\nu},$ with $\nu > 0.$
%The tunneling time thus goes as 
%$\tau_{\Gamma} \sim \exp{(2d_{ab}/Gamma_{\nu})}$. 

%%%%%%%%%%%%%%%%%%%%%%%%%%%%%%%%%%%%%%%%
%\vspace{0.4cm} 

%\noindent(ii) {\textit{Random TSP}}

%\vspace{0.45cm}
%%%%%%%%%%%%%%%%%%%%%%%%%%%%%%%%%%%%%%%%%
%%%%%\subsubsection{Random TSP}
\subsubsection{The Traveling Salesman Problem}
\begin{figure}
\centering
\includegraphics*[width=5.5cm]{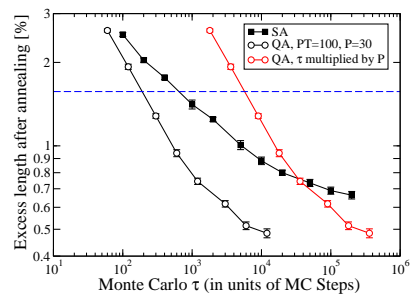}
\caption{\footnotesize{
  Average residual excess length found after CA and QA for a
  total time $\tau$ (in MC steps), for the $N=1002$ instance \texttt{pr1002} 
of the TSPLIB.The dashed horizontal line represents the best out of
         1000 runs of the Lin-Kernighan algorithm.
QA is clearly faster than CA
 (taken from Battaglia et al 2005).}}
\label{QTSP}
\end{figure}

\noindent
Quantum annealing of TSP with a random metric (i.e., the distance
$d_{ij}$ between the $i$-th and the $j$-th city is chosen randomly)
in infinite dimension 
using PIMC
was also found to be more efficient than CA in finding an approximately
minimal tour (Marto\u{n}\'{a}k et al 2004). An $N$-city tour in a 
random TSP problem can be represented by a configuration of $N^{2}$
constrained Ising spins, and the tour length, i.e., the $\mathcal{H}_{c}$, 
by the Hamiltonian 
(\ref{TSP-Ham}). Now to do the annealing, one needs to introduce a set of moves
(spin flip operations) that satisfies the constraints. Classically one very
important class of moves is the 2-opt moves, which starting from a valid tour,
can generate all possible tours without generating any invalid link. Let a
valid tour contains two links $i\rightarrow j$ and $k\rightarrow l$. A 2-opt
move may consist of removing those two links and establishing the following
links: $i\rightarrow k$ and $j\rightarrow l$ (here $i$ denotes $i$-th city).
Classical annealing of the Hamiltonian can be done by restricting the
Monte Carlo moves within 2-opt family only. However, for the quantum case,
one needs to design a special transverse field (non-commuting spin-flip term)
to enforce the constraints. It can be realized by a spin-flip term of the
form $S^{+}_{<k,i>}S^{+}_{<l,j>}S^{-}_{<j,i>}S^{-}_{<l,k>}$, where 
the operator $S^{-}_{i,j}$ flips down ($ +1 \rightarrow -1$) 
the Ising spins $S^{z}_{ij}$ and $S^{z}_{ji}$ when they are in $+1$ state, and
similarly for the flip-up operators $S^{+}_{i,j}$. However, to avoid 
the Suzuki-Trotter mapping with these complicated kinetic terms, 
a relatively simple
kinetic term of the form 
$\mathcal{H}_{kin} = -\Gamma(t)\sum_{<i,j>} \left( S^{+}_{<i,j>} 
+ H. C.\right)$ is used for the quantum to
classical mapping. The Monte Carlo moves were kept restricted within the
2-opt family to avoid invalid tours. The results were tested on an instance of
printed circuit board with $N = 1002$. PIMC-QA was seen to do 
better than the CA and also much better than standard Lin-Kernighan 
algorithm, as
shown in Fig. \ref{QTSP}.
%PIMC-QA works much better than CA and also Lin-Kernighan algorithm
%in finding the least path for the printed circuit board version of TSP, as
%shown in Fig. \ref{QTSP}.
The relaxation behavior of residual path-length for
TSP (see Fig. \ref{QTSP}) is found to be quite similar 
to that of the 2-d EA glass discussed earlier (see Battaglia et al 2005).
This indicates that a random TSP also has spin glass-like 
potential energy landscape (PEL), 
as has already been 
hinted by the replica analysis study of the problem 
discussed in an earlier section (Sec. IIC.2).
In that case the Landau-Zener 
tunneling picture would also be applicable for TSP,
and little improvement can be expected by following the 
real-time Schr\"{o}dinger 
dynamics instead of Monte Carlo methods.

There are cases where PIMC-QA is not as
successful as CA. An example (discussed below) of such a problem 
belongs to the class of K-Satisfiability problem (or K-SAT). 
In a $K$-SAT problem,
there is a
given Boolean function,
which is the Boolean sum (connected by logical OR operations) 
of a number of clauses, 
each of which is a Boolean
 product (connected by logical AND operations) of $K$ Boolean variables 
(binary variables taking values 0 or 1 only)
taken random from a given set of $p$ variables (same variable may occur 
in various clauses simultaneously). Given
such a function, the task is to
find an assignment for the Boolean variables for which the number of
violated clauses [the clauses assigned with the $0$ value] is the the minimum.
The studies on this class of problems is extensive
and it even includes the connection between its hardness and
satisfiability-unsatisfiability phase transition 
(Monasson et al 1999). Remarkable progress has been made in formulating faster 
algorithms for solving it (see Mezard et al 1987,  
Hartmann and Rieger 2002) based on these understandings. 

But in the case of the random 3-SAT 
problem (K-SAT problem with K= 3)
using linear schedule for decreasing $\Gamma$, 
PIMC-QA gives much worse results than CA (Battaglia et al 2004).
Both CA and PIMC-QA
are worse than ad-hoc local search heuristics like WALKSAT.
In the case of application of PIMC-QA in image restoration, on the other hand,
the performance is exactly the same as that of CA (Inoue 2005).
Even for
a particle in a simple double-well potential, it seems that 
with naive Monte Carlo
moves, PIMC-QA can produce results which are much worse than that one could
obtain from real-time Schr\"{o}dinger evolution of the system. There is
in fact no general prescription to choose the right moves that will 
do the job,
unless one has a precise idea about the PEL of the problem. 
The choice of the kinetic term
to be introduced into the problem is somewhat arbitrary but the 
performance of the algorithm depends on crucially on that.
%The knowledge of the phase diagram of the system may provide a
%reasonable guideline for the choice.
It has been found that a relativistic kinetic term can do a better
job than a non-relativistic one in the case of a particle in
a double-well potential 
(Battaglia et al 2005). 
PIMC-QA also 
suffers from difficulty in calculating the Suzuki-Trotter equivalent of the
quantum Hamiltonian with arbitrary kinetic term designed to satisfy the
constraints of the problem
(Marto{n}'{a}k et al, 2004). 
The constraints may be taken care of while
making Monte Carlo moves, but that may not always produce expected results.
Finally, presence of finite temperature in the problem does not allow one to
focus exclusively on the role of  the quantum fluctuations 
in the problem. Above
all, like any other Monte Carlo method, PIMC-QA is going to do worse if the
number of reasonably good approximate solutions are only few,
and there is no overall 
gradient in the landscape to guide towards them.

%%%%%%%%%%%%%%%%%%%%%%%%%%%%%%%%%%%%%%%%%%%%%%%%%%%%%%%%%%%%%
%\vspace{0.4cm} 
%
%\noindent(iii) {\textit{Random Field Ising Model: How Choice of
%Kinetic Term Improves Annealing Results}}
%
%\vspace{0.45cm}
%%%%%%%%%%%%%%%%%%%%%%%%%%%%%%%%%%%%%%%%%%%%%%%%%%%%%%%%%%%%%%%%%
\subsubsection{Random Field Ising Model: 
How a Choice of Kinetic Term Improves Annealing Results}
\noindent
QA algorithms enjoy an extra flexibility which a thermal annealing
algorithm cannot. A QA algorithm can have a host of choices for 
its kinetic term. A good choice can 
bring about a lot of betterments. This point is
illustrated nicely by Morita and Nishimori (see Morita and Nishimori 2007a)
for QA of random field Ising model, by introducing a ferro-magnetic transverse
field interaction, in addition to the
conventional single-spin-flip transverse field term (as present in 
the Hamiltonian (\ref{SK-Q})).

The Hamiltonian of the random-field Ising model with the
standard single-spin-flip transverse term
is given by

\begin{equation}
\mathcal{H}(t) = \mathcal{H}_{C} + \mathcal{H}_{Kin}^{(1)},
\label{RFIM}
\end{equation}
\noindent where
\begin{equation}
\mathcal{H}_{C} = 
-J\sum_{<ij>}^{N}S_{i}^{z}S_{j}^{z} -\sum_{i=1}^{N} h_{i}^{z}S_{i}^{z},
\label{Hpot}
\end{equation}
\noindent
$h_{i}^{z}$ being the on site random field assuming values $+1$ or $-1$ 
with equal probabilities and
$<ij>$ denotes sum over the nearest 
neighbor on a 2-dimensional square lattice, and
%%%%%%%%% corr-ref-3-1 %%%%%%%%%%%%%%%
\begin{equation}
\mathcal{H}_{Kin}^{(1)} = \Gamma(t) \sum_{i=1}^{n} S_{i}^{x}.
\end{equation}
\noindent
The result of QA in such a system in not 
satisfactory when  $J$ is much larger than 
$h_{i}^{z}$ (Sarjala et al 2006). 
If a ferromagnetic transverse term of the
form 
\begin{equation}
\mathcal{H}_{kin}^{(2)} = -\Gamma(t)\sum_{<ij>}^{N}S_{i}^{x}S_{j}^{x}
\label{Kin-2}
\end{equation}
\noindent
is added to the Hamiltonian (\ref{RFIM}), 
the result of QA is seen to improve considerably
(Morita and Nishimori 2007a). This happens (as indicated by
exact diagonalization results on small system) because
the ferromagnetic transverse field term effectively 
increases the gap between the ground state and
the first excited state and thus decreases
the characteristic timescale for the system. 
This is an example of how one can utilize the
flexibility in choosing the kinetic term in QA to 
formulate faster algorithms. This also indicates how the
knowledge of the phase diagram of the system, the position of the
quantum critical point in particular (where the gap tends to vanish),
helps in choosing additional kinetic terms and thus allows for the
annealing paths that can avoid, to some extent, 
the regions of very low gap. 
 
%\subsubsection{Zero Temperatute Quantum Monte Carlo Annealing}
%
%NOT DECIDED YET!!!

\subsection{Quantum Annealing Using Real-time Adiabatic Evolution}
\noindent
%Following a real-time Schr\"{o}dinger evolution is not easy in a 
%classical computer, but might be a tractable one in quantum computer. Thus
%quantum annealing scheme provides a ground for continuous time
%analog quantum computation. 
QA is basically the analog version of quantum computation. Like the
conventional analog quantum computation, the hardware realization of
adiabatic quantum annealing is rather problem specific. 
But once realized, it 
follows the real-time Schr\"{o}dinger evolution, 
whose exact
simulation is always intractable 
(run time grows exponentially with the system size)
 for classical computers and often
also even for digital quantum 
computers (see Nielsen and Chuang 2000). The annealing
behavior with the real-time Schr\"{o}dinger evolution is hence
an important issue and may show features distinctly different
from any Monte Carlo annealing discussed so far.

The first analog algorithm in this line was due to
Farhi and Gutmann (1998) for solving Grover's search problem. 
Grover (1997)
showed that quantum mechanical search can reduce an $\mathcal{O}(N)$ 
classical search time to $\mathcal{O}(\sqrt{N})$ time in finding out a marked
item form an unstructured data-base. 
In the analog version, the problem was to use quantum evolution to 
find a given marked state among  $N$ orthonormal states.
The algorithm was formulated in the following way. There were
$N$ mutually orthogonal normalized states, 
the $i$-th state being denoted by $|i\rangle$. Among all of them,
only one, say, the $w$-th one, has energy $E\ne0$ and the rest all have zero
energy. Thus the state $|w\rangle$ is ``marked'' energetically and the system
can distinguish it thus. Now the question is how fast the system can evolve
under a certain Hamiltonian (our oracle) 
so that starting from an equal superposition of the $N$ states one reaches
$|w\rangle$. It was shown (Farhi and Gutmann 1998) 
that by evolving the system under a
time-independent Hamiltonian of the
form
\begin{equation}
\mathcal{H}_{tot} = E|w\rangle\langle w| + E|s\rangle\langle s|,
\label{Grover-ana-1}
\end{equation} 
\noindent
%%%%%%%%% corr-ref-3-1 %%%%%%%%%%%%%%%
where 
\begin{equation}
|s\rangle = \frac{1}{\sqrt{N}}\sum_{i=1}^{N} |i\rangle,
\end{equation} 
\noindent
no improvement over Grover's $\sqrt{N}$ speedup is possible. 
Later the problem was recast in the form of a spatial search 
(Childs and Goldstone 2004), 
where there
is a $d$-dimensional lattice and the basis state $|i\rangle$ 
is localized at the $i$-th lattice site. Similarly, as 
before, the on-site potential energy $E$ is zero everywhere except
at $|w\rangle$, where it is 1.
%%%with on-site potential energy $E$ equal to
%%%zero on every site, except one marked one, say, the $w$-th one, 
%%%where $E = 1$. We take position basis vectors ($|i\rangle$ denoting a state 
%%%localized at the $i$-th site), as the entries of the data-base.
The objective is to reach the marked state starting from the
equal superposition of all the $|i\rangle$s. The kinetic term is 
formulated through the Laplacian of the lattice, 
which effectively introduces uniform hopping
to all nearest neighbors from any given lattice site and is kept
constant. The model is in essence an
Anderson model (Santoro and Tosatti, 2006) with only a single-site disorder of 
strength $\mathcal{O}(1)$. Grover's speed up can be achieved for 
$d\ge4$ with such a Hamiltonian and no further
betterment is possible. The algorithms succeeds only near the
critical value of the kinetic term 
(i.e., at the quantum phase transition point).
An adiabatic quantum evolution algorithm
for Grover search was formulated in terms of an orthonormal
complete set of $l$-bit Ising-like basis vectors, where the potential energy
of a given basis vector (among the $2^{l}$ ones) is 1, and for rest of all, it
is 0. The kinetic term is just the sum of all single bit-flip terms
(as the transverse-field term in Eq. (\ref{SK-Q})). It 
connects each basis vector to all those that can be
reached from it by a single bit flip. 
The kinetic term is reduced from a
high value to zero following a linear schedule. 
A detailed analysis in light of the adiabatic theorem
showed that one cannot even retrieve Grover's 
speedup sticking to the global adiabatic 
condition with fixed evolution rate as given by Eq. (\ref{adia-cond}). 
This is so
because the minimum value of the gap goes 
as $\Delta_{min} \sim 2.2^{-l/2}$
(Farhi et al 2000),
and the spin-flip kinetic term being local, the numerator 
$|\langle\dot{\mathcal{H}_{tot}}\rangle|$
of the adiabatic factor (see Eq. (\ref{adia-cond})) is at best 
$\mathcal{O}(l).$ However, the Grover's speedup can be recovered
if the condition of adiabaticity is maintained 
locally at every instant of the evolution and the rate
is accelerated accordingly whenever possible (Roland and Cerf, 2001).
%There is an equivalence between adiabatic quantum computation and
%the gate-based standard quantum computation (see Mizel et al, 2007, 
%Santoro and Tosatti 2007)
%in the sense that for any task that can be performed by a standard quantum 
%circuit, there exists a time-dependent Hamiltonian $\mathcal{H}(t)$
%that does the same job evolving adiabatically.  
Adiabatic QA following real-time Schr\"{o}dinger evolution 
for satisfiability
problems 
also gives strikingly different result (for smaller system-size, though)
compared to that obtained using PIMC-QA.
Adiabatic QA of an NP-complete problem, namely, the exact cover problem 
(as described below) 
is studied for small 
systems, where a quadratic system-size dependence was observed 
(Farhi et al 2001). 
In this problem also, the basis vectors are 
the complete set of $2^{l}$ orthonormal $l$-bit basis vectors,
denoted by $\{|z_1\rangle\|z_2\rangle...|z_l\rangle\}$, each $z_i$ is
either 0 or 1. 
The problem consists of a cost function $\mathcal{H}_C$ which is the
sum of many 3-bit clause functions $h_{Cl}(z_{i},z_{j},z_{k})$, each
acting on
arbitrarily chosen bits $z_{i}, z_{j}$ and $z_{k}$. 
The clause function is such that
$h_{Cl}(z_{i},z_{j},z_{k}) = 0$ 
if the clause $Cl$ is satisfied by the states
$|z_i\rangle$, $|z_j\rangle$ and $|z_k\rangle$ 
of the three bits, or else $h_{Cl}(z_{i},z_{j},z_{k}) = 1$.
The cost Hamiltonian is given by $\mathcal{H}_{C} = \sum_{Cl} h_{Cl}$. 
Thus if a basis state $|\{z_i\}\rangle$ dissatisfies $p$ clauses, then
$\mathcal{H}_{C}|\{z_i\}\rangle = p|\{z_i\}\rangle$. The question is whether
there exists a basis vector that satisfies 
all the clauses for a given $\mathcal{H}_{C}$. There may be many basis
vectors satisfying a clause. All of them will be the
ground state of $H_{C}$ with zero eigenvalue. If the ground state
has a non-zero (must be a positive integer then) 
eigenvalue, then it represents the basis
with lowest number of violated clauses, the number
being given by the eigenvalue itself. The total Hamiltonian
is given by
\begin{equation}
\mathcal{H}_{tot}(t) = \left(1 - \frac{t}{\tau}\right)\mathcal{H}_{kin} 
+ \frac{t}{\tau} \mathcal{H}_{C}, 
\label{QA-Ham}
\end{equation}
\noindent
where $\mathcal{H}_{kin}$ is again the sum of 
all the single bit-flip operators.
%connects any given basis state with all those
%basis vectors who are reachable from it 
%by flipping of a single bit. 
The initial state at $t=0$ is taken to be
the ground state of $\mathcal{H}_{kin}$, which is an equal superposition
of all basis vectors. The system is then 
evolved according to the time-dependent
Schr\"{o}dinger equation up to $t = \tau$. The value of $\tau$ required to
achieve a pre-assigned success probability are noted for different system sizes.
The result showed a smooth quadratic system size dependence for $l \le 20$ 
(Farhi et al 2001). 
The result is quite encouraging (since it seems to give a polynomial
time solution for an NP-complete problem for small system size), 
but does not really assure a quadratic behavior
in the asymptotic ($l\rightarrow\infty$) limit.

A quadratic relaxation behavior
($\epsilon_{res} \sim 1/\tau^{\zeta} $, $\zeta \sim 2$) was 
also reported in (Suzuki and Okada 2005a, 2005b) for real-time
adiabatic QA (employing exact method for small systems and 
DMRG technique for larger
systems) of a one-dimensional tight-bonding model with random site
energies and also for a random field Ising
model on 2-d square lattice. In the first case, the kinetic term was due to 
hopping between nearest neighbors, while in the second case it was simply
the sum of single-spin flip operators as given in Eq. (\ref{SK-Q}).  
Wave function annealing results using similar DMRG technique for spin glass on
ladder has been reported by Rodriguez-Laguna (2007).  

It has been demonstrated that for finite-ranged systems, 
where the interaction
energy can be written as a sum of interaction energies involving few 
variables, quantum adiabatic annealing and thermal annealing may not differ
much in efficiency. However, for problems with spiky  
(very high but narrow) barriers in the PEL 
(which must include infinite-range terms
in the Hamiltonian, since in finite-range systems, barrier heights
can grow at best linearly with barrier width), quantum annealing 
does much better than CA (Farhi and Goldstone 2002) 
as has been argued by Ray et al
(Ray et al 1989). Quantum adiabatic evolution has been employed earlier
to study the SK spin glass in a transverse field; stationary characteristics
over a range of the transverse field are calculated by varying
the transverse field adiabatically (Lancaster and Ritort 1997).
 
As pointed out earlier in the case of PIMC-QA, the kinetic term plays a 
crucial role in adiabatic QA also. This is mainly because the gap 
$\Delta(\Gamma)$
between the ground state and first excited state depends on the form of
$\mathcal{H}_{kin}$. For certain choice of $\mathcal{H}_{kin}$, it might become
very small for some $\Gamma$. One then can switch on some
additional kinetic term (adjusted 
by another parameter $\Gamma^{\prime}$) to 
 increase the dimensionality of the phase diagram
 and avoid the small gap zone by
taking a suitable by-pass. In fact such a strategy can convert a failure into
success (Farhi et al 2002).

So far it has been argued that the advantage of quantum annealing 
derives
mainly from the fact that quantum tunneling can penetrate
through very high but narrow barriers, which are very
hard to jump over thermally. In this way the 
cost-configuration landscapes more accessible to local moves. 
This is a key feature that works even in the case of
quantum Monte Carlo methods like PIMC-QA, where one
finally samples only a very small section of the
configuration space. However, another remarkable advantage
that quantum mechanics provides is the ability to ``sense'' the
whole configuration space simultaneously through a delocalized
wave function. This sensing is largely handicapped by presence
of random disorder in the system, because the wave function
tends to localize in many places, often being unable to 
 to pick up the deepest well very distinctly. But this
feature may be utilized in searching a
golf-course-like PEL, namely a flat landscape with
very deep and narrow wells occurring very rarely. If there is only one
deep well, the problem is just the spatial version of the Grover's problem.
As we have seen, if the depth $\chi$ of the well is $\mathcal{O}(1)$ 
(independent of $N$),
then no more than $\mathcal{O}(\sqrt{N})$ speed-up can be achieved. This can be
interpreted as the inability of an 
exceedingly large system to sense a given small
wound. So one might want to make the depth of the well large enough so that
it cannot be scaled away as the system-size is increased.

\begin{figure}
%%\resizebox{7.0cm}{!}
%%{\rotatebox{270}{\includegraphics{Chi=N-tau-Vs-N-min-srvrd.ps}}}
%%\resizebox{7.0cm}{!}
%%{\rotatebox{0}{\includegraphics{ConstG-Chi=N=1E6-tau-Vs-Pkw.ps}}}
\resizebox{7.0cm}{!}{\rotatebox{0}{\includegraphics{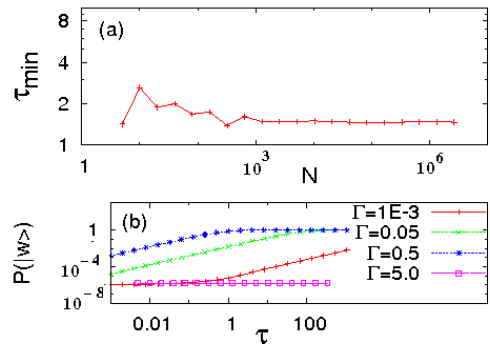}}}
\caption{\footnotesize{Panel (a) in the figure 
shows numerical verification of the 
$N$-independence of the minimum time $\tau_{min}$
to achieve success probability $P(|w\rangle)=0.33$.  
Initially, the system is delocalized
equally over all sites and evolves with time
according to time dependent Schr\"{o}dinger equation
with the Hamiltonian (\ref{ad-ham}).
As expected from exact analytical result for the adiabaticity condition, 
it is seen that $\tau_{min}$ becomes
independent of $N$. 
Panel (b) shows the variation of final probability $P(|w\rangle)$ of finding
the state $|w\rangle$ with annealing time $\tau$ for different final value of $\Gamma$, for 
$N = 10^6$.}}
\label{qan}
\end{figure}

 We first consider the case
where the well-depth goes as $\chi \sim \alpha N$, where $\alpha$ 
is some constant. 
For a well depth of such order for a spatial search at 
infinite dimension (kinetic term is infinite-ranged) one gets
a stunning result: The evolution time that guarantees any given probability of
success becomes independent of $N$ (Das 2007; see Fig. \ref{qan}). 
However, if $\chi$ scales like 
$\chi \sim N^{\gamma}$, with $\gamma > 1$, then the speed-up is lost again - a
consequence of quantum-mechanical non-adiabaticity.
We formulate the problem in the following way. As in the case of a
spatial search discussed above, we denote the state 
localized at the $i$-th cite by $|i\rangle$. All the cites except $|w\rangle$
has zero on-site potential energy, while $|w\rangle$ have an on-site potential
well of depth $\chi(t)$. The system is embedded in infinite dimensions and 
thus there is a uniform tunneling term $\Gamma$ between any
two sites. To do quantum annealing, we evolve $\chi(t)$ from zero to its
final value $\chi_{0}$ keeping $\Gamma$ fixed at some moderate constant value.  
The total Hamiltonian is given by 
%$\mathcal{H}_{tot} = \mathcal{H}_c + \mathcal{H}$
%with $\mathcal{H}_C = -chi_{0}|w\rangle\langlew|$ and $\mathcal{H}_{kin} = $ 

\begin{equation}
\mathcal{H}_{tot} = -\chi_{0}\left(1-t/\tau\right)|w\rangle\langle w| 
- \Gamma\sum_{i,j; i\ne j}|i\rangle\langle j|.
\label{ad-ham}
\end{equation} 
 
The time-dependent eigen problem can be solved for the above 
Hamiltonian.
It can be shown exactly that if 
$\chi_{0} \sim \alpha N$, then in the $N\rightarrow\infty$
limit, the adiabatic factor
\begin{equation}
\frac{ 
|\langle \dot{\mathcal{H}_{tot}} 
\rangle|_{max}}{\Delta^2_{min}(t)} \sim \frac{\alpha}{4\Gamma^{2}},
\label{ad-adia}
\end{equation}
\noindent 
which means that the run time becomes independent 
of $N$, and the ground state has $\mathcal{O}(1)$ overlap with
the target state $|w\rangle$. We also confirm it numerically by
evaluating $\tau_{min}$, which is the minimal $\tau$ 
required for obtaining a success
probability $P(|w\rangle) = 0.33$ for different $N$ through many 
decades. 
%In the figure we have shown how $\tau_{min}$ varies with the number of sites $N$.
Here we have chosen a moderate constant $\Gamma$, 
and have evolved the well depth $\chi$ with
time. 
The evolution is computed by solving the
time-dependent Schr\"{o}dinger equation numerically and 
 $\tau_{min}$ is obtained with an accuracy of $10^{-4}$ 
by employing a 
bisection scheme.
The results (Fig. \ref{qan}a) 
clearly show that $P(|w\rangle)$ tends to becomes 
independent of $N$ as $N$ becomes larger and larger. This is
completely in accordance with the analytical result (see Eq. (\ref{ad-adia})). 

The relaxation behavior for large $N$ for a given annealing time $\tau$
depends on the value of $\Gamma$ (see Fig. \ref{qan}b). 
If $\Gamma$ is too small,
the system takes a longer time to feel the changes in the landscape, 
and hence the adiabatic 
relaxation requires a longer time (the adiabatic factor becomes bigger; 
see Eq. (\ref{ad-adia})). On the
other hand, if $\Gamma$ is too large, the ground state itself is 
pretty delocalized, 
and hence the final state, though closer to the ground state, 
has again a small overlap with
the target state $|w\rangle$. For $\Gamma = 0.5$, the result is best. 
For higher and lower values
of $\Gamma$, the results are worse, as shown in Fig. \ref{qan}(b). 
The relaxation behavior is 
seen to be linear with the annealing time $\tau$.

\subsection{Annealing of a Kinetically Constrained System}
\noindent
The
adiabatic theorem of quantum annealing assures convergence of a quantum
algorithm when one starts with the initial (trivial)
ground state of the Hamiltonian and evolves slowly 
enough so that the system is
always in the ground state of the instantaneous Hamiltonian. However, the
benefit of tunneling may be extended even in cases where one does not
precisely know the eigenstate of the initial Hamiltonian (say for a given
unknown PEL) and hence is unable to start with it. One
might rather start with a wave-packet (a superposition of many eigenstates
of the Hamiltonian) 
that explores the potential energy
landscape. Quantum tunneling will still allow it to move more easily through
the PEL than a classical particle if the landscape has
many high but narrow barriers. A semi-classical treatment of such a
non-stationary annealing has been discussed in the
context of a kinetically constrained
system (KCS) (Das et al 2005).

We demonstrate here the effectiveness of quantum annealing
\index{quantum annealing} in the context
of a certain generalized Kinetically Constrained Systems (KCS) (Fredrickson and Andersen et al 1984).
KCSs are simple model systems having trivial ground state 
structures and static properties, but a complex 
relaxation behavior due to some explicit constraints 
introduced in the dynamics of the system. 
These systems are very important in understanding how much of
the slow and complex relaxation behavior of a glass can be attributed to its
constrained dynamics alone, leaving aside any complexity of its energy 
landscape structure. In KCSs one can view the constraints to be 
represented by infinitely high energy barriers appearing dynamically.
%%%%%%%%%%%%%%%%%%%%%%%%%%%%%%%%%%%%%%%%%%%%%%%%%%%%%%%%%%%%%%%%%%%%%%% 
%To study annealing, we generalize\index{kinetic constraints!generalized} 
%such models by 
%allowing the height of such 
%kinetically occurring barriers to be finite but large. 
%%%%%%%%%%%%%%%%%%%%%%%%%%%%%%%%%%%%%%%%%%%%%%%%%%%%%%%%%%%%%%%%%%%%%%%%%

We discuss quantum annealing in the context of a KCS,
\index{kinetic constraints!generalized}
 which can be represented by a generalized version of 
the East model\index{East model!generalized}
(Jackle and Eisinger 1991); a one dimensional KCS. 
We also compare the results with 
that of thermal annealing done in the same system.    
The original East model\index{East model}
 is basically a one-dimensional chain of 
non-interacting classical Ising (`up-down') spins in a longitudinal field
$h$, say, in downward direction. The ground state of such 
a system is, trivially, all spins down. 
A kinetic constraint\index{kinetic constraints} is introduced in the model by
putting the restriction that the $i$-th spin 
cannot flip if the $(i-1)$-th spin
is down. Such a kinetic constraint\index{kinetic constraints}
 essentially changes the topology of
the configuration space, since the shortest path between any two 
configurations differing by one or more forbidden flips is
increased in a complicated manner owing to the blockage of the `straight'
path consisting of direct flips of the dissimilar spins. 
Further, the constraint becomes 
more limiting as more spins turn down, as happens in the late approach to 
equilibrium. As a result, the
relaxation processes have to follow more complex and 
lengthier paths, giving
rise to exponentially large timescale ($\sim e^{1/T^2}$; 
Jackle and Eisinger 1991). 

%\subsection{Model}
%\noindent
In the Das model (Das et al 2005) there is a chain of
asymmetric double-wells (each with infinite boundary walls), 
each having a particle localized within them. The asymmetry is due to an
energy difference of $2h$ between the two wells of a double well.
The particle in one of the two (asymmetric) wells can change its location
to the other well stochastically, either due to the thermal fluctuations or
the quantum fluctuations present in the system. The generalized kinetic
constraint\index{kinetic constraints}
 is introduced by assuming that if the particle in the ($i-1$)-th
double-well resides in the lower one of the two wells, 
then there appears a barrier
of height $\chi$ and width $a$ between the two wells of 
the $i$-th double-well. In such a situation the particle in the $i$-th
double-well has to cross the barrier in order to change its location from
one well to the other. 
%(Fig. \ref{Wells}(b)).%%%%%%%%%%%%%%% 
On the other hand, if the particle 
of the $(i-1)$-th is in its upper well, there is no 
such barrier to cross for the particle to go from one well to the other.
%% (Fig. \ref{Wells}(a)).%%%%%%%%%%%%%%%%%%%%%%%% 
Following the approximate mapping done in case of a
symmetric double-well (Chakrabarti et al 1996),  
this model can be approximately represented by a generalized version
of the East model\index{East model}, 
where each Ising spin is in a local longitudinal field $h$ in
the downward direction. The spin at the $i$-th site sees a barrier of height 
$\chi$ and width $a$ between its two energy states  
when the $(i-1)$-th spin is down, 
%%%%%(Fig. \ref{Wells}(b)), %%%%%%%%%%%%%%%
while no such barrier occurs
for the $i$-th spin when the $(i-1)$-th spin in up. 
%%%%%(Fig. \ref{Wells}(a)).%%%%%%%%%%%%%% 
This kinetic constraint\index{kinetic constraints} is the same
in both cases irrespective of whether the dynamics is classical or quantum.

%\begin{figure}
%\begin{center}
%\resizebox{5.0cm}{!}{\includegraphics{East1a.eps}}
%\end{center}
%\caption{\small{Potential energy wells for the spin at  
%site $i$, when ($i-1$)-th
%spin is (a) up and (b) down, with the external field $h$ in the downward
%direction and barrier height $\chi$ very large and the width $a$ small.
% For the classical 
%generalized East model\index{East model!generalized}, 
%flipping across the barrier in (b) is a thermal
%jump (at any finite $T$). In the quantum model considered here,
% probability for crossing the barrier in (b) is due to quantum tunnelling
% through it at a finite $\Gamma$.}}
%\label{Wells}
%\end{figure}

When the dynamics of the particle is due to quantum 
fluctuations, the tunneling
probabilities come from the following semi-classical picture of
scattering of a particle in a double-well with
infinitely remote outer boundaries. 
%($a\rightarrow \infty$ in Fig. \ref{Wells}).
If a particle is put in one of the wells of such a double-well with some
kinetic energy (actually the expectation value) $\Gamma$, then it will
eventually be scattered by the 
separator (a barrier or step) between the two wells. 
In such a scattering,
there is a finite probability $P$ that the particle manages 
to go to the other
well. We calculate $P$ from the simple picture of scatterings of 
a particle by one dimensional potentials as described below. 
In the thermal case we take simple Boltzmann probabilities for
crossing the same barriers.
The minimum of the energy of the Ising chain (equivalent to 
the potential energy of the chain of the double-wells) trivially
corresponds to the state
with all the spins down, i.e., aligned along the longitudinal field $h$ 
(where all the particles are in their respective lower wells). 
To reach the ground state in the quantum case, we start with a 
very large initial value of $\Gamma$ and 
then reduce it following an exponential schedule given by 
$ \Gamma = \Gamma_0\exp{(-t/\tau_Q)}.$ Here $t$ denotes the time and 
$\tau_Q$ sets the effective time scale of annealing. 
At zero-temperature the slow spin flip dynamics
occurs only due to the tunneling (kinetic energy) term
$\Gamma$ and hence the system ceases to have any relaxation dynamics in the 
limit $\Gamma\rightarrow 0 $. The barriers are characterized by
a height $\chi$ and a width $a$, the barrier area being $g = \chi\times a$.
%%It may be mentioned here that
%%in absence of any analytical expression for the tunnelling probability
%%in asymmetric case of the type discussed here (see e.g., (Cordes and Das, 2001)), 
%%we employ
%%the asymmetric barrier tunnelling probabilities available. 
%A:Murphy}. 
Similarly, in thermal case, we start with a high initial temperature
$T_0$ and reduce it eventually 
following an exponentially decreasing temperature schedule given by
$ T = T_0\exp{(-t/{\tau_C})}$; 
 $\tau_C$ being the time constant for the
thermal annealing schedule.
 Here, when the $(i-1)$-th spin is down, the flipping probability for the $i$-th
spin ($\sim \exp{(-\chi/T)}$).
 Otherwise, it flips with probability $P = 1$ if it were in the up state and 
 with Boltzmann probability $P = \exp{(-h/T)}$ if it were in the down state.
%%%%%%%%% corr-ref-3-11 %%%%%%%%%%%%%%%%%%%
Here in the quantum case, the probability of crossing the barrier depends on
$g$ so that the barrier-width $a$ plays a role, while in the thermal case,
only $\chi$ sets the crossing timescale irrespective of $a$. 

%%\subsection{Simulation and Results}  
%%\noindent 
In the simulation (Das et al 2005), $N$ Ising spins
($S_{i} = \pm 1,\quad i = 1,..., N $) were taken on a linear chain with
periodic boundary condition. 
The initial spin configuration is taken to be random, so that the
magnetization $m = (1/N)\sum_{i}S_{i}$ is practically negligible
($m_i \approx 0$).
 We then start
with a tunneling field $\Gamma_0$ and follow the 
zero-temperature (semi-classical)
Monte Carlo scheme as mentioned above, using the spin flip probabilities $P$'s
appropriate for the four cases I-IV. 
Each complete run over the entire lattice
is taken as one time unit and as time progresses, $\Gamma$ is decreased from
its initial value $\Gamma_{0}$ according to 
$\Gamma = \Gamma_{0}e^{-t/\tau_{Q}}$. 
\begin{figure}
%\resizebox{12.0cm}{!}{\rotatebox{270}{\includegraphics{EastC_10L.eps}}}
\resizebox{7.5cm}{!}{\includegraphics{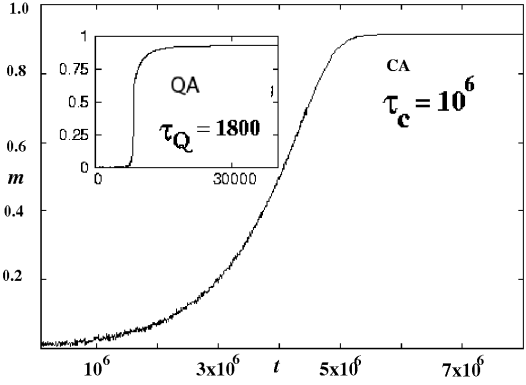}}
\caption{\footnotesize{Comparison between classical and quantum
annealing for a chain of $5\times 10^4$ spins (for the same initial disordered
configuration with $m_i \sim 10^{-3}$). We show the results for
 $\tau_Q = 1.8\times 10^{2}$ (for quantum) and 
$\tau_C = 10^{6}$ (for classical) with $h=1$; a lower
$\tau_C$ would not produce substantial annealing.  
Starting from the same initial values $\Gamma_{0} = T_{0} = 100$, 
(and $g =100$ in the quantum case) 
we observe that classical annealing requires about
$10^{7}$ steps, whereas quantum annealing
\index{quantum annealing!of a kinetically constrained system}
 takes about $10^{4}$ steps
for achieving the same final order $m_f\sim 0.92$.}}
\label{Q-cl-comp}
\end{figure} 
The results of thermal and 
quantum annealing\index{quantum annealing!of a kinetically constrained system}
 are compared in Fig. \ref{Q-cl-comp}
for the same order of initial value and time constant
 for $\Gamma$ and $T$ (the barrier height $\chi$ is 
taken to be $1000$ in both the cases, and
$g$ was taken to be $100$ in the 
quantum annealing\index{quantum annealing!of a kinetically constrained system}
 case,
 or equivalently the barrier width $a$ is taken to be
of the order of $0.1$).
It is observed that to achieve a similar degree of
annealing (attaining a certain final magnetization $m_f$),
 starting from the same disordered configuration, one typically requires much
smaller $\tau_Q$ compared to $\tau_C$; typically, 
$\tau_C \sim 10^{3}\times\tau_Q $ for equivalent annealing 
(for similar optimal values of final order $m_f \sim 0.92$). For annealing
with final order $m_f \sim 1$, we find $\tau_C \sim 10^{4}\times\tau_Q$. This
comparison depends on the barrier characteristics (the value of $g$).

\subsection{Experimental Realization of Quantum Annealing}
\begin{figure}
\resizebox{6.5cm}{!}{\rotatebox{270}{\includegraphics{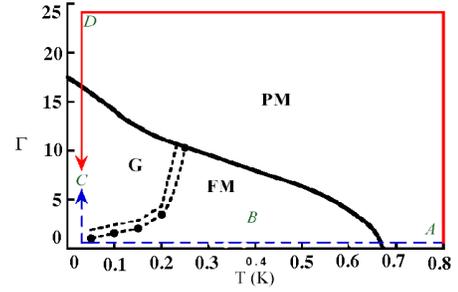}}}
\caption{{\footnotesize{
Experimental realization of QA and CA in
LiHo$_{0.44}$Y$_{0.56}$F$_4$ is illustrated on its 
phase diagram on
the temperature
$T$ and transverse field $\Gamma$ 
(measured by the magnitude of the external laboratory
field in kOe) plane.
The material behaves like a conventional ferromagnet in the region labeled FM,
and shows slow relaxation in the glassy domain wall state labeled G.
The two paths of relaxations, from an initial point A to a final point C 
on the phase diagram are shown by arrow-headed lines.
Along the classical path A$\rightarrow$B$\rightarrow$C (dashed arrow) the
transverse field is not applied until the end, so that the relaxations observed
are purely thermal. Whereas, along the quantum path 
A$\rightarrow$D$\rightarrow$C (continuous arrow), there is a segment
where the temperature is small enough and the transverse field is high, so 
that the fluctuations are mainly quantum mechanical. 
Relaxations observed along the quantum path are often found to be much  
faster than those observed along the classical path at low enough
temperature. 
 Taken from Aeppli and Rosenbaum (2005).}}} 
\label{QA-Exp}
\end{figure}
\noindent
Brooke et al (1999) showed experimentally
(see also Aeppli and Rosenbaum 2005) that 
the relaxation behavior in reaching deep inside the glass
phase in Fig. \ref{QA-Exp} depends on the path chosen. 

%In particular,
%the relaxation along the 
%classical annealing path of reducing the temperature 
%$T$ at a fixed rate (at
%zero transverse field) is much 
%3 to 30 times (depending on the temperature and the %) 
%slower in reaching a given point deep
%into the glass phase than that following a quantum annealing path, where
%the transverse field is reduced in same rate keeping the
%(low) temperature fixed.
A TISG, realized by a sample of LiHo$_{0.44}$Y$_{0.56}$F$_{4}$ in
laboratory transverse field ($\Gamma$), was taken from a
high temperature paramagnetic to a low temperature glassy phase following
two separate paths in the $\Gamma-T$ plane (see Fig. \ref{QA-Exp}). 
Along the classical path of cooling (CA), the
transverse field was kept zero throughout, 
and was switched on only after reaching the final temperature. The
quantum cooling (QA), on the other hand, was done in the
 presence of
a high transverse field, 
which was lowered only on reaching the final temperature. 
As the sample is cooled, 
spectroscopy of the sample at different temperatures 
(both during CA and QA) was done to 
reveal the nature of the distribution of spin
relaxation time scales.
The QA produced states whose relaxation was up to
30 times faster than those produced by the CA 
at low temperature (see Aeppli and Rosenbaum 2005). This
clearly indicates that quantum tunneling is much more effective
in exploring the configuration space
in the glassy phase than thermal jumps (as indicated
in Fig. \ref{QA-SA}).
An experimental realization of quantum adiabatic annealing for
3-bit instances of MAXCUT problem using NMR technique 
has also been reported (see Steffen et al 2003). Here, the 
the smoothly varying time dependent Hamiltonian 
was realized by the technique of quantum simulation 
(see Nielsen and Chuang 2000), where a smooth time 
evolution was achieved approximately (Trotter approximation)
through the application of a series of discrete unitary
operations. The results indicate existence of an optimal run time 
of the algorithm. 
  
%%%%%%%%%%%%%%%%%%%%%%%%%%%%%%%%%%%%%%%%%%%%%%%%%%%%%
\section{Convergence of Quantum Annealing Algorithms}
%%%%%%%%%%%%%%%%%%%%%%%%%%%%%%%%%%%%%%%%%%%%%%%%%%%%%
\noindent
Here we briefly summarize some important recent results
derived by Morita and Nishimori (2006, 2007b) on the convergence of QA
 algorithms for TIM systems. 
%%%(Morita and Nishimori, 2006; Morita and Nishimori, 2007b). 
The results are valid for both the
Quantum Monte Carlo and the real-time Schr\"{o}dinger evolution
versions of QA.

%In order to seek the convergence criterion for quantum Monte Carlo versions
%of the QA algorithms, one needs to analyse the
%inhomogeneous Markov Chain employed to simulate the evolution of the system
%with under a time-dependent Hamiltonian. 
%The Hamiltonian
%here is the same as given in (\ref{SK-Q}) with a time dependence
%in the transverse field $\Gamma = \Gamma(t)$. No assumption regarding either 
%on the nature
%of distribution of $J_{ij}$ or the spatial dimensionality is required.  
%As required for any QA algorithm, $\Gamma$ is kept initially 
%much higher than
%the interaction term, and gradually decreased to zero following some
%annealing schedule $\Gamma(t)$. The Monte Carlo simulation of the
%system is run throughout the schedule, and one has to find out what is 
%the fastest schedule that ensures convergence of the final state to the
%true ground state of the system. In order to do that one has to find the
%minimal criterion for strong ergodicity of the inhomegeneous Markov Process
%that is employed to simulate the system. A Markov process is said to be 
%strongly ergodic if it always converges finally to a unique stationary
%distribution irrespective of its initial states.

The Hamiltonian
here is the same as given in (\ref{SK-Q}) with a time dependence
in the transverse field $\Gamma = \Gamma(t)$. No assumption regarding either
the nature
of the distribution of $J_{ij}$ or the spatial dimensionality is required. 
%%%%%%%%%%%%%%%%%%%%%%%%%%%%%%%%%%%%%%%%%%%%%%%%%%%%%%%%%%%%%%%%%%%%%%%
%One may represent the
%partition function in a somewhat more general form
%%instead of taking the Suzuki-Trotter partition function for
%%the TIM Hamiltonian, 
%%(as given in Eq. \ref{Hameff}),%%%%%%%%%%%%% 
%%one may deal with a more general partition function of the form
%%\begin{equation}
%Z(t) = \sum_{x\in \mathcal{C}} \exp{\left(-\frac{F_{0}(x)}{T_{0}} 
%-\frac{F_{1}(x)}{T_{1}}\right)},
%\label{Partition}
%\end{equation}
%\noindent where $x$ denotes a configuration of the system (Ising spin here)
%and $\mathcal{C}$ denotes the set of all configurations. Here 
%$F_{0}(x)$ is the cost function (spin-spin interaction part 
%of the Hamiltonian here) 
%and $F_{1}(x)$ is the term derived from the non-commuting quantum 
%kinetic energy part (transverse field term in TIM). The strength of
%the tarnsverse field is tuned through the parameter $T_1$. 
%%%%%%%%%%%%%%%%%%%%%%%%%%%%%%%%%%%%%%%%%%%%%%%%%%%%%%%%%%%%%%%%%%%%%%%%%%%
In order to perform QA at temperature $T$ using PIMC, one
constructs the Suzuki-Trotter equivalent (see Appendix 1) $d+1$ 
dimensional classical system of the
 $d$ dimensional quantum system,
and the resulting (classical) system is simulated using a suitable
inhomogeneous Markov Chain. The transverse field 
$\Gamma(t)$ is tuned from a high value
to zero in course of simulation. 
It can be shown that at the end of the simulation the final 
distribution converges to the ground state of the classical part of the
Hamiltonian irrespective of the initial distribution 
(strong ergodicity) if
\begin{equation}
\Gamma(t) \ge MT\tanh^{-1}\left[{\frac{1}{(t+2)^{2/RL}}}\right],
\label{TIM-SE}
\end{equation} 
\noindent
where $M$ is the number of Trotter replicas in the $d+1$ dimensional
Suzuki-Trotter equivalent system. Here $R$ and $L$ are constants
depending on the system size $N$, the spin-flip dynamics appointed for 
the simulation, the temperature 
$T$, etc. For large $t$ 
the bound reduces to
\begin{equation}
\Gamma(t) \ge MT \frac{1}{(t+2)^{2/RL}}.
\label{Gamma-rlx-2}
\end{equation}
It is remarkable that
in contrast to the inverse
logarithmic  decay of temperature required for convergence
of CA (see Eq. \ref{SA-conv}), QA requires only a power-law decay of
the transverse field. In this sense, quantum annealing is much faster
than the classical annealing for TIM. 
Similar result can be derived
for more general form of Hamiltonian (see Morita and Nishimori 2006).  
However, the advantage gained here does not change the complexity class
of an NP-complete problem, since $RL$ is of the order of $N$ and hence the
convergence time is thus still exponential in $N$. 

For real-time Schr\"{o}dinger dynamics at $T = 0$, the bound for the
decay of the transverse field is again of the form
\begin{equation}
\Gamma(t) \ge (\xi t)^{-\frac{1}{(2N-1)}},
\label{TDSE-conv-1}
\end{equation} 
\noindent 
where $\xi$ is exponentially small in $N$ for large $N$. Though the dynamics
of PIMC-QA and real-time Schr\"{o}dinger evolution are completely different,
the power-law bound on the annealing schedule is strikingly similar. 

%%%%%%%%%%%%%%%%%%%%%%%%%%%%%%%%%%%%%%%%%%%%%%%%%%%%%%%%%%%%%%%%%%%%%%%%%%%
\section{Quantum Quenching}
%%%%%%%%%%%%%%%%%%%%%%%%%%%%%%%%%%%%%%%%%%%%%%%%%%%%%%%%%%%%%%%%%%%%%%%%%%%
\noindent
Instead of annealing, one can quench a systems by reducing the 
transverse field from $\Gamma_{i} (> \Gamma_c)$ to 
 $\Gamma_{f} (< \Gamma_c)$ and follow the relaxation dynamics after that.
Here $\Gamma_c$ denotes the quantum critical
point for the system.
This can help to prepare a state of the system with substantial 
order in a very short time.
Recently, quantum quenching dynamics in different systems, in particular
for sudden quenching 
across a quantum critical point, have been studied extensively
(Sengupta et al 2005,  
 Calabrese and Cardy 2006,
Das et al 2006). The (exact) results for
the after-quench dynamics, following a sudden quench
of an infinite-range Ising model in transverse field
(Das et al 2006, see Appendix 2) is shown in Fig. \ref{semi-classical}. 
The results
indicate that 
%%%%%%%%%%%%%%%%%%%%%%%%%%%%%%%%%%%%%%%%%%%%%%%%%%%%%%%%%%%%%%%%%%%%%%%
%Here we discuss the 
%quenching dynamics in an infinite-range interacting Ising ferromagnet
%in transverse field (see Appendix D for exact analysis).
%%%%%%%%%%%%%%%%%%%%%%%%%%%%%%%%%%%%%%%%%%%%%%%%%%%%%%%%%%%%%%%%%%%%%
for a purely quantum system (small $S$), 
one might reach a non-stationary oscillatory state with a substantial
value for the
long-time average (over the largest timescale) of the order
$\langle (\sum S_{i}^{z})^{2}(t) \rangle$
if the quenching is done from $\Gamma_i > \Gamma_c$ to $\Gamma_f = \Gamma_c/2.$
Thus one gets a state 
(though non-stationary) which has considerable order, 
in just one step quenching. However, if the quenching is carried out with a
high but finite rate, one ends up reaching an excited state with
topological defects (see Kibble 1976, Zurek 1985, Sengupta et al 2008).  

%%%%%%%%%%%%%%%%%%%%%%%%%%%%%%%%%%%%%%%%%%%%%%%%%%%%%%%%%%%%%%%%%%%%%%

\begin{figure}[h]
\resizebox{5.5cm}{!}{\includegraphics{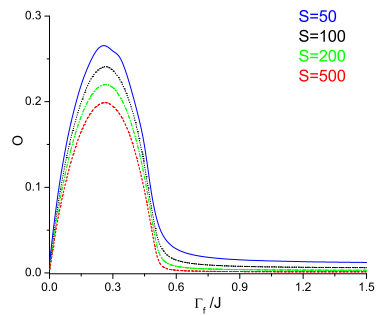}}
\caption{\footnotesize{Plot of the long-time average (see Appendix 2) 
$O = \langle (S_{tot}^{z})^{2} \rangle$
%\quad S(t) \equiv \sum_{i} S^{z}_{i}$ 
as a function of $\Gamma_f/J$ for
different $S$. In the plot the solid (blue), dotted (black), dash-dotted
(green) and the dashed (red) lines represents respectively the results for
$S = 50$, $S = 100$, $S = 200$ and $S = 500$ (color online). 
$O$ peaks around $\Gamma_f/J=
0.25$ and the peak value decreases 
with increasing $S$. For all plots we have
chosen $\Gamma_i/J=2$. (The figure is taken from Das et al 2006).}}
\label{semi-classical}
\end{figure}
\section{Summary and Discussions}
\noindent
Unlike the 
gate-based quantum computers (see, e.g.,
Ekert and Jozsa 1996, Nielsen and Chuang 2000,
Galindo and Martin-Delgado 2002), the
annealing of a physical system towards the optimal state (encoded in the
ground state of the final Hamiltonian) in the classical limit naturally
achieves analog quantum computation. As discussed here, the
utilization of the
quantum mechanical tunneling through the classically localized states in 
annealing of glasses
has opened up this new paradigm for 
analog quantum computation of hard optimization problems through
adiabatic reduction of the quantum fluctuations.

We reviewed here the recent success in annealing, or optimizing the
cost functions of complex systems, utilizing quantum fluctuations rather than
the thermal fluctuations (see Santoro and Tosatti 2006 for a more
technical review). As mentioned already, following the early indication in
Ray et al (1989) and the pioneering demonstrations,
theoretically by
Amara et al (1993), Finnila et al (1994),
Kadowaki and Nishimori (1998), Farhi et al (2001) and Santoro
et al (2002), and experimentally by Brooke et al  
(1999), the quantum annealing technique has now emerged as 
a successful technique for optimization of complex cost functions. 
The literature, exploring its success and also its limitations, is also
considerably developed at present. 
%%has been reviewed here. 

These are introduced here through discussions of the 
mapping of such hard problems to classical spin glass problems, 
discussions on
quantum spin glasses, and consequent annealing.
\noindent
The
physics of classical spin glasses (see Sec. II)
offers us the
%%(introduced in 1975 and not completely explored yet)
%has already contributed enormously to our 
knowledge of the
of the energy or the thermodynamic potentials and 
their landscape structures.
%%of the
%%unusually slow (glassy) dynamics of many-body
%%systems in presence of frustration and disorders. 
Mapping of computationally
hard problems like the TSP etc problems
to their corresponding classical spin glass models 
also helped understanding their
complexity (Sec. IIC).
%%%%%%%%% corr-r2-3 %%%%%%%%%%%%%%%%%%%%%%%%%%%%%%
The timescale for tunneling through an energy 
barrier quantum mechanically, involves not only the height of the
barrier, but also its width; the narrower the barrier, the 
faster the
tunneling (for constant height). Thermal 
fluctuations, on the other hand, see only the barrier height for crossing it.
This reduction of tunneling time with the barrier width
helps relaxing a quantum system much faster in some cases.
This leads to the quantum annealin, a
framework for constructing
general heuristics
to find approximate solutions of hard optimization
problems. While simulated annealing employs
the strategy of slow cooling, physical or in simulations,
to find the ground state of glassy systems,
quantum annealing employs quantum
fluctuations (see Sec. III).
As mentioned before, this effectively reduces Planck's
constant to zero to reach the classically optimized (minimum cost) state.
This reduction, when
done completely adiabatically, guarantees that
 the system will be found
in the ground state of the classical glass at the end (provided there is no
crossing of energy levels with the ground state in the course of
evolution, and one has started initially with the ground state of
the Hamiltonian). This has already been realized 
experimentally (see Sec. IIID), where faster relaxation %%% corr-r2-4 %%%%%%
towards the ground state is achieved  %%% corr-r2-4 %%%%%%
by reducing the external field (inducing changes in the
tunneling field), rather than by reducing the temperature. 
In this way 
analog quantum computation is  
realized
through a novel route. Recently an equivalence between
the adiabatic QA and standard gate-based quantum 
computation has also been establised
(Aharonov et al 2007, Mizel et al 2007). 
\begin{acknowledgments}
We are grateful to A. Chakrabarti, 
P. Ray, D. Sen, K. Sengupta 
and R. B. Stinchcombe for their collaborations
at different stages of this work.
We acknowledge G. Aeppli, A. Dutta,
H. Nishimori, G. Santoro, P. Sen and E. Tosatti for useful
discussions. 
%We also thank M. Maity for scrutinizing
%the final manuscript.
\end{acknowledgments}

\section{Appendix}

\subsubsection{Suzuki-Trotter Formalism}

%This is a formalism where the partition function of a $d$ dimensional
%quantum system at temperature $T$ is mapped to that of an
%equivalent $d+1$ dimensional classical
%system. 
\noindent
Here we
illustrate this formalism by applying it to a TIM having a Hamiltonian
%We start with Transverse Ising Hamiltonian
%\begin{eqnarray}
\begin{equation}
\mathcal{H} = -\Gamma\sum_{i=1}^N S_{i}^{x} - \sum_{(i,j)}J_{ij}
S_{i}^{z}S_{j}^{z} 
 \equiv \mathcal{H}_{kin} + \mathcal{H}_{C}.
\label{ST-QHam}
\end{equation}
%\end{eqnarray}

\noindent The canonical partition function of $\mathcal{H}$ reads
%%%%%%%%% corr-ref-3-1 %%%%%%%%%%%%%%%
\begin{equation}
 Z = Tr e^{-(\mathcal{H}_{kin}+\mathcal{H}_{C})/T}. 
\end{equation}

\noindent Now we apply the Trotter formula
%%%%%%%%% corr-ref-3-1 %%%%%%%%%%%%%%%
\begin{equation} 
\exp{(A_1 + A_2)} = \lim_{M\rightarrow \infty}
\left[ \exp{A_{1}/M}\exp{A_{2}/M} \right]^M, 
\end{equation}
\noindent even when $[A_{1},A_{2}] \ne 0$. On application of this, $Z$ reads
\begin{eqnarray}
 Z &=& \lim_{M\rightarrow \infty}\sum_{i}
\langle s_i |\left[ \exp{(-\mathcal{H}}_{kin}/MT)\right.\times\nonumber \\
&& \left.\exp{(-\mathcal{H}_{C}/MT)} \right]^M |s_i \rangle. 
\end{eqnarray}

\noindent Here $s_i$ represent the $i$-th spin configuration of the whole
system, and the above summation runs over all such possible configurations
denoted by $i$. Now we introduce $M$ number of identity operators
%%%%%%%%%%%%%%%%%%%%%% Identity operator %%%%%%%%%%%%%%%%%%%%%%%%%%%%
%$$\mathcal{I} = \sum_{i}^{2^N}|s_{i,k}\rangle\langle s_{i,k}|,\qquad
% k = 1,2,...M.$$
%%%%%%%%%%%%%%%%%%%%%%%%%%%%%%%%%%%%%%%%%%%%%%%%%%%%%%%%%%%%%%%%%%%%
in between the product of $M$ exponentials in $Z$, and have
\begin{eqnarray} 
Z &=& \lim_{M\rightarrow\infty} Tr \prod_{k=1}^{M}
\langle S_{1,k}... S_{N,k}|\exp{\left(\frac{-\mathcal{H}_{kin}}{MT}
  \right)}\times \nonumber\\
& & \exp{\left(\frac{-\mathcal{H}_{C}}{MT}\right)}|
S_{1,k+1} ...S_{N,k+1}\rangle,\nonumber
\end{eqnarray}
\noindent and periodic boundary condition would imply $S_{N+1,p} =
S_{1,p}$.
Now,
$$\prod_{k=1}^{M} \langle S_{1,k}...S_{N,k}|
\exp{\left(\frac{1}{MT}\sum_{i,j}J_{ij} S_{i}^{z} S_{j}^{z} \right)}|
S_{1,k+1}...S_{N,k+1}\rangle$$
 $$= \exp{\left[\sum_{i,j=1}^{N} \sum_{k=1}^{M}
\frac{J_{ij}}{MT} S_{i,k}S_{j,k} \right]},  $$
where $S_{i,k} = \pm 1$ are the eigenvalues of $S^z$ operator 
(see Hatano and Suzuki, 2005), and
\begin{eqnarray}
& &\prod_{k=1}^{M} \langle S_{1,k}...S_{N,k}|
\exp{\left[\frac{\Gamma}{MT}\sum_{i}
S_{i}^{x}\right]}|S_{1,k+1}...S_{N,k+1} \rangle \nonumber \\
&=& \left(\frac{1}{2}\sinh{\left[ \frac{2\Gamma}{MT} \right]}
 \right)^{\frac{NM}{2}}\times \nonumber\\
& & \exp{ \left[\frac{1}{2}
\ln {\coth{\left(\frac{\Gamma}{MT}
\right)}}\sum_{i=1}^{N} \sum_{k=1}^{M}S_{i,k}S_{i,k+1} \right]},\nonumber
\end{eqnarray}
\noindent
giving the effective classical Hamiltonian (\ref{E-32}), equivalent to the
quantum one in (\ref{ST-QHam}).
In the above equation 
$M$ should be at the order of $1/T$
 ($\hbar= 1$) for a meaningful comparison
of the interaction in the Trotter direction with that in the original
Hamiltonian. For $T\rightarrow 0$, $M\rightarrow\infty$, and
the Hamiltonian represents a
system of spins in a ($d$+1)-dimensional lattice, because of 
%which is
%one dimension higher than the original $d$-dimensional Hamiltonian,
%as is evident from
the appearance of one extra label $k$ for each spin variable.
Thus corresponding to each single quantum spin variable $S_i$
in the original Hamiltonian we have an array of $M$ number of classical
replica spins $S_{i k}$.
This new (time-like)
dimension along which these classical spins are spaced is known as
Trotter dimension.

\subsubsection{Quantum Quenching of a Long Range TIM}
%Recently, quantum quenching dynamics in different systems, in particular
%for quenches across a quantum critical point, is being studied extensively
%(cardy et all 2007, Das et al 2006). Here we discuss the
%exact quenching dynamics in an infinite-range interacting Ising ferromagnet
%in transverse field (Das et al 2006).
\noindent
Let us consider a system of spin-$\frac{1}{2}$ objects governed by the
Hamiltonian
%\begin{equation}
$\mathcal{H} = -\frac{J}{N}\sum_{i>j}^{N} S_{i}^{z}S_{j}^{z}
              -\Gamma\sum_{i}^{N}S_{i}^{x}.$
%\label{Q-qnch-Ham-1}
%\end{equation}
%\noindent 
It can easily be rewritten as
\begin{equation}
\mathcal{H} = -\frac{J}{N}\left(S_{tot}^{z}\right)^{2}
              -\Gamma S_{tot}^{x},
\label{Q-qnch-Ham-2}
\end{equation}
\noindent where $S_{tot}^{z} = \sum_{i} S_{i}^{z}$,
$S_{tot}^{x} = \sum_{i} S_{i}^{x}$ and a constant
$(J/2N)\sum_{i}(S_{i}^{z})^2 = J/8$ from $\mathcal{H}$ in (\ref{Q-qnch-Ham-2}).
%Within the mean field approximation, the
The above Hamiltonian can again be cast into the simplified form
%\begin{equation}
$\mathcal{H} = \vec{h}.\vec{S_{tot}},$
%\label{Q-qnch-Ham-3}
%\end{equation}
%\noindent 
where $\vec{h} = Jm \hat{z} - \Gamma \hat{x}$, giving
the mean field (exact in this long-range limit) equation
%\begin{equation}
$m \equiv
\langle S_{tot}^{z}\rangle =  
\frac{|\vec{h}.\hat{z}|}{|\vec{h}|}
\tanh{\left(\frac{|\vec{h}|}{T}\right)} 
= \frac{Jm}{2\sqrt{\Gamma^{2} + J^{2}m^{2}}},$
%\label{Q-qnch-mag}
%\end{equation}
\noindent
at $T = 0$
and $\hat{z}$, $\hat{x}$ denote unit vectors along $z$ and $x$ directions
respectively.
This gives $m = 0$ for $\Gamma > \Gamma_{c}$ and $m\ne 0$ for
$\Gamma < \Gamma_{c} = J/2.$
Since the model is infinite-ranged one, the mean field
approximation becomes exact and one can readily express $\vec{S}$ in terms of
its polar components as $\vec{S} = S(\sin{\theta}\cos{\phi},
\sin{\theta}\sin{\phi}, \cos{\theta})$, $S$ being the total angular momentum.
One can immediately utilize the classical equation of motion
%\begin{equation}
$\frac{d\vec{S}}{dT} = \vec{S}\times\vec{h}.$
%\label{dSdt}
%\end{equation}
Considering the above
equation for the $z$ and $x$ components, we get
%%(Eqs. \ref{Q-qnch-Ham-3} and \ref{dSdt})
%%\begin{eqnarray}
\begin{equation}
\frac{d\theta}{dt} = \Gamma\sin{\theta}\quad \mathrm{and} \quad
\frac{d\phi}{dt} = -\frac{J}{2}\cos{\theta} + \Gamma\cot{\theta}\cos{\phi}.
\label{dSdt}
\end{equation}
%%\end{eqnarray}
\noindent
Here we have $S = N/2$. If the system is now quenched from above
 its quantum critical point $\Gamma > \Gamma_{c},$
%% \quad \Gamma_{c} = J/2$,
finally to a
$\Gamma_{f} < \Gamma_{c},$
%%(i.e., from $\langle S^{z} \rangle \equiv m = 0$ to $m \ne 0$),
 then one can write (see Das et al 2006), equating the energies of the states
with and without any order respectively,
%%using (\ref{Q-qnch-Ham-2}),
%\begin{equation}
$\Gamma_{f} = \frac{J}{4}\cos^2{\theta} + \Gamma_{f}\sin{\theta}\cos{\phi}.$
%\label{Gamma-f}
%\end{equation}
%\noindent
Using this, one gets from Eq. (\ref{dSdt})
\begin{equation}
\frac{d\theta}{dt} = \frac{\sqrt{\Gamma_{f}^2\sin^2{\theta} -
[\Gamma_{f} - J/4\cos^2{\theta}]^{2}}}{\sin{\theta}} \equiv f(\theta).
\label{dThetadt}
\end{equation}
\noindent
This has zeros (turning points) at $\theta_{1} =
\sin^{-1}{(|1 - 4\Gamma_{f}/J|)}$ and $\theta_{2} = \pi/2$. One can
therefore obtain $\langle (S^{z}_{tot})^{2}\rangle =
\langle \cos^{2}{\theta}\rangle =
\mathcal{N}/\mathcal{D} $, where $\mathcal{N} =
\int_{\theta_{1}}^{\theta_{2}} d\theta \cos^2{\theta}/f(\theta) =
4\sqrt{8\Gamma_{f}(J - 2\Gamma_{f})/J}$ and $\mathcal{D} =
\int_{\theta_{1}}^{\theta_{2}} d\theta/f(\theta)$, giving a behavior
shown in Fig. \ref{semi-classical}.

%%%%%%%%%%%%%%%%%%%%% BIBLIOGRAPHY%%%%%%%%%%%%%%%%%%%%%%%%%%%%%%%%%%%%%


\begin{thebibliography}{99}

\bibitem{Aeppli-chapt} Aeppli G. and Rosenbaum T. F. (2005), in 
Das and Chakrabarti (2005), pp 159-169 
%%(2005)

\bibitem{Aharonov}Aharonov D., van Dam W., Kempe J., Landau Z., Lloyd S. and Regev O. (2007),
SIAM Journal of Computing, {\bf 37} 166
%arXiv:quant-ph/0405098 (2004)

\bibitem{Alvarez-ritort} Alvarez J. A. and Ritort F. (1996),
J. Phys. A {\bf 29}, 7355

\bibitem{Amara} Amara P., Hsu D., and Straub J. E. (1993), 
J. Phys. Chem. {\bf 97} 6715

\bibitem{Apolloni-89}
Apolloni B., Carvalho C. and de Falco D. (1989),
Stochastic Processes and their Applications {\bf 33} 233

\bibitem{Apolloni-90}
Apolloni B., Cesa-Bianchi N., and de Falco D. (1990),
in Albeverio et al (Eds.) {\it Stochastic Processes, Physics and Geometry},
World Scientific, pg. 97

\bibitem{Barahona} Barahona F. (1982), 
J. Phys. A {\bf 15} 3241 
%(1982)
%11

\bibitem{3-Sat} Battaglia D. A., Santoro G. E. and Tosatti E. (2004), 
Phys. Rev. E
{\bf 70} 066707 
%%(2004).%28

\bibitem{Santoro-chapt} Battaglia D. A., Stella L., Zagordi O., 
Santoro G. E. and Tosatti E. (2005), in Das and Chakrabarti (2005),
pp 171-204 
%(2005).
% {\textit{Quantumam Annealing and Related
%Optimization Methods}}, A. Das and B. K. Chakrabarti (Eds.),
%Lecture Note in Physics, {\bf 679},
%Springer-Verlag, Heildelberg (2005)%30

\bibitem{TSP-Hamersley} Bearwood J., Halton J. H., Hammersley J. H. (1959), 
Proc. Camb. Phil. Soc. {\bf 55} 299 
%%(1959) 

\bibitem{BC:Bhatt} Bhatt R. N. (1998), in
\textit{Spin Glasses and Random Fields},  Young A. P. (Ed.),
World Scientific, Singapore, pp. 225 - 249 
%%(1998)%16

\bibitem{Binder} Binder K. and Young A. P. (1986), 
Rev. Mod. Phys. {\bf 58} 801 
%(1986)%1 

\bibitem{Bray} Bray A. J. and Moore M. A. (1984), J. Phys. C Lett.
{\bf 17} L463

\bibitem{Brook}Brooke J., Bitko D., Rosenbaum T. F. and Aeppli G. (1999),
Science {\bf 284} 779

\bibitem{Brook-2} Brooke J., Rosenbaum T. F., Aeppli G. (2001), 
Nature {\bf 413} 610

%\bibitem{Calabrese} Calabrese P. and Cardy J. (2007), arXiv:0704.1880 
\bibitem{Cardy-2006} Calabrese P. and Cardy J. (2006),
Phys. Rev. Letts. {\bf 96} 136801 

\bibitem{BKC:Ishi} Chakrabarti B. K. (1981), Phys. Rev. B {\bf{24}} 4062 
%%(1981)

\bibitem{BC:BKC} Chakrabarti B. K., Dutta A. and Sen P. (1996),
 {\textit{Quantum Ising Phases and Transitions in Transverse Ising Models}},
Springer-Verlag, Heidelberg 
%%(1996)

%% see also S. Sachdev,
%%{\textit {Quantum Phase Transitions}}, Cambridge Univ. Press,
%%Cambridge (1999)%17

\bibitem{Anirban-TSP} Chakraborti A. and Chakrabarti B. K. (2000),
Euro. Phys. J. B {\bf 16} 667
%arXiv:cond-mat/0001069
%(2000)

\bibitem{Childs} Childs A. M. and Goldstone J. (2004), 
arXiv:quant-ph/0306054  
%(2004)%33

%\bibitem{A:Cordes} Cordes J. G. and Das A. K., Superlatt.
%Microstr., {\bf 29} 121 (2001)

\bibitem{ad-pap} Das A. (2007), In preparation; 
see also the thesis Das A. (2008), Jadavpur Univ., Kolkata
%(2007)%38

\bibitem{BC:adbk}
Das A. and Chakrabarti B. K. (2005), Eds., 
{\textit{Quantum Annealing and Related
Optimization Methods}}, Lecture Note in Physics, {\bf 679}, 
Springer-Verlag, Heidelberg 
%%(2005)%19


\bibitem{AD-kcs} Das A., Chakrabarti B. K. and Stinchcombe R. B. (2005),
Phys. Rev. E {\bf 72} 026701 
%%(2005)%25

\bibitem{AD-quench} Das A., Sengupta K., Sen D., Chakrabarti B. K. (2006),
Phys. Rev B {\bf 74} 144423 
%%(2006)

\bibitem{Dotsenko} Dotsenko V. (2001), {\textit{Introduction to the 
Replica Theory of Disordered Statistical Systems}}, Cambridge University Press,
Cambridge, UK 
%%(2001)%3

\bibitem{E-A} Edwards S. F. and Anderson P. W. (1975), 
J. Phys. F: Met. Phys. {\bf 5} 965
%(1975)

\bibitem{Ekert} Ekert A. and Jozsa R. (1996), 
Rev. Mod. Phys. {\bf 68} 733 

\bibitem{Farhi-Grover} Farhi E. and Gutmann S. (1998), 
Phys. Rev. A {\bf 57} 2403
%(1998)%31 

\bibitem{Farhi-arXiv-bst} Farhi E., Goldstone J., Gutmann S. and
Sipser M. (2000), Preprint quant-ph/0001106 
%v1 
%(2000)%35

\bibitem{Farhi} Farhi E., Goldstone J., Gutmann S., 
 Lapan J., Ludgren A. and
Preda D. (2001), Science {\bf 292} 472 
%(2001)%24 

\bibitem{Farhi-comp} Farhi E. and  Goldstone J (2002), 
arXiv:quant-ph/0201031 
%v1 
%(2002)%36

\bibitem{Farhi-path} Farhi E, Goldstone J. 
and Guttman S. (2001), arXiv:quant-ph/0208135  
%(2002)%37

\bibitem{Finnila} Finnila A. B., Gomez M. A.,
Sebenik C., Stenson C., and Doll D. J. (1994), Chem. Phys. Lett. {\bf 219}, 
343 
%(1994)%22   

\bibitem{Fisher} Fisher D. S. and Huse D. A. (1986), Phys. Rev. Lett.
{\bf 56}, 1601

\bibitem{A:Fred} Fredrickson G. H. and Andersen H. C. (1984),
Phys. Rev. Lett. {\bf 53} 1224 
%(1984)
% G. H. Fredrickson and H. C. Andersen,
%J. Chem. Phys. {\bf 83} 5822 (1985)

\bibitem{Fu} Fu Y. and Anderson P. W. (1986), J. Phys. A {\bf 19} 1620 
%(1986)%10

\bibitem{Galindo} Galindo A. and Martin-Delgado M. A. (2002),
Rev. Mod. Phys. {\bf 74} 347 

\bibitem{Garey} Garey M. R., Johnson D. S. (1979), 
{\it Computers and Intractability: Guide to the theory of NP-Completeness}
Freeman; San Francisco 
%(1979)

\bibitem{Gaviro} Gaviro P. S., Ruiz-Lorenzo J. J. and Tarancon A.
(2006), J. Phys A {\bf 39} 8567 

\bibitem{Geman} Geman S. and Geman D. (1984), IEEE Trans. 
Pattern Anal. Mach. Intell. {\bf 6} 721 
%(1984)%9

\bibitem{Goldschmidt} Goldschmidt Y. Y. and Lai P.-Y. (1990), Phys. Rev. Lett.
{\bf 64} 2467 
%(1990)

\bibitem{Grover} Grover L. K. (1997), Phys. Rev. Lett. {\bf 79} 325 
%(1997)%32

\bibitem{BC:Guo} Guo M., Bhatt R. N. and Huse D. A. (1994),
Phys. Rev. Lett. {\bf{72}} 4137 
%(1994) 
%D. Fisher, Phys. Rev. B {\bf 50} 3799
%(1994)%14

\bibitem{Heiko} Hartmann A. and  Rieger H. 2002, 
{\it Optimization in Physics}, Wiley VCH, Darmstadt
%Rieger H. (2005a), in Das and Chakrabarti (2005),

\bibitem{Hatano} Hatano N. and Suzuki M. (2005), in Das and Chakrabarti (2005),
pp - 37-67 
%(2005)

\bibitem{Huse} Huse D. A and Fisher D. S. (1986), Phys. Rev. Lett.
{\bf 57} 2203 
%(1986)%27

\bibitem{Inoue} Inoue J.-I. (2005), in Das and Chakrabarti (2005), 
pp 259-296 
%(2005)
% {\textit{Quantumam Annealing and Related
%Optimization Methods}}, A. Das and B. K. Chakrabarti (Eds.),
%Lecture Note in Physics, {\bf 679},
%Springer-Verlag, Heildelberg (2005)%29

\bibitem{BC:Ishi}
Ishii H. and Yamamoto T. (1985), J. Phys. C {\bf{18}} 6225 
%%(1985)%12

\bibitem{A:Jackle} Jackle J. and Eisinger S. (1991), Z. Phys. B {\bf 84} 115 
%%(1991)
%;M. A. Munoz, A. Gabrielli, H. Inaoka and L. Peitronero, Phys. Rev. E {\bf 57}
% 4354 bitem{(Fredrickson and Andersen et al 1984) G. H. Fredrickson and H. C. Andersen,
%Phys. Rev. Lett. {\bf 53} 1224 (1984); G. H. Fredrickson and H. C. Andersen,
%J. Chem. Phys. {\bf 83} 5822 (1985)(1998); P. Sollich, M. R. Evans, Phys. Rev. Lett. {\bf 83} 3238,
% F. Ritort and P. Sollich, Adv. Phys., {\bf 52} 219 (2003)

\bibitem{BC:Kado} Kadowaki T. and Nishimori H. (1998),
Phys. Rev. E {\bf 58} 5355 
%%(1998)%21

\bibitem{Kibble}  Kibble T. W. B., (1976) J. Phys. A {\bf 9} 1387

\bibitem{BC:Kim} Kim D.-H. and Kim J.-J. (2002), Phys. Rev. B {\bf 66} 054432 
%%(2002);

\bibitem{BC:Kirk} Kirkpatrick S., Gelatt C.D. 
and Vecchi M. P. (1983), Science,
{\bf{220}} 671
%%(1983)%8

\bibitem{Krzakala} Krzakala F., Houdayer J., Marinari E., Martin O. C.
and Parisi G. (2001), Phys. Rev. Lett. {\bf 87} 197204

\bibitem{Lancaster-Ritort} Lancaster D. and Ritort F (1997),
J. Phys. A {\bf 30}, L41

\bibitem{Marinari} Marinari E., Parisi G., Ruiz-Lorenzo J.J. and
Zuliani F. (1998), Phys. Rev. Lett. {\bf 82} 5176

\bibitem{Martonak} Marto\u{n}\'{a}k R.,
G. E. Santoro and E. Tosatti (2002), Phys. Rev. B {\bf 66} 094203 
%%%(2002)%23 

\bibitem{Santoro-TSP} Marto\u{n}\'{a}k R., 
Santoro G. E. and Tosatti E. (2004), Phys. Rev. E {\bf 70} 057701 
%%%(2004)%26


%Mezard M., Parisi G., Sourlas N., Toulouse G. and
%Virasoro M., J. Physq {\bf 45} 843 (1984)

\bibitem{(BC:Parisi)} Mezard M., Parisi G. and Virasoro M. A. (1987),
{\it Spin Glass Theory and Beyond} World Scientific Lect. Note in Phys. 
{\bf 9}, Singapore and references therein%4

\bibitem{Miller} Miller J. and Huse D. A. (1993),
Phys. Rev. Lett {\bf 70} 3147

\bibitem{Mizel} Mizel A., Lidar D. A. and Mitchell M. (2007), 
Phys. Rev. Lett. {\bf 99} 070502 

\bibitem{Monasson} Monasson R., Zecchina R., Kirkpatrick S., Selman B. and
Troyansky L. (1999), Nature {\bf 400} 133

\bibitem{Moore} Moore M. A., Bokil H. and Drossel B. (1998),
Phys. Rev. Lett. {\bf 81} 4252 

\bibitem{Nishi-conv} Morita S. and Nishimori H. (2006), 
J. Phys. A {\bf 39} 13903 
%%%(2006)%38

\bibitem{Nishi-Random-field}Morita S. and Nishimori H. (2007a),
J. Phys. Soc. Jap. {\bf 76} 06400
%arXiv:quant-ph/0702214 
%%%(2007a)

\bibitem{Nishi-conv-2}Morita S. and Nishimori H. (2007b),
arXiv:quant-ph/0702252 
%%%(2007b)

\bibitem{Nielsen} Nielsen M. A. and Chuang I. L. (2000), 
{\textit Quantum Computation and Quantum Information}, 
Cambridge University Press, Cambridge

\bibitem{BC:Nishi} 
Nishimori H. (2001), {\textit{Statistical Physics of Spin Glasses
and Information Processing: an Introduction,}} Oxford University Press, Oxford
%(2001)%2

%\bibitem{Papa-2} Papadimitriou C. H.
%{\it Computational Complexity},
%Addison-Wesley Publishing Company, New York (1994)%6

\bibitem{Papa} Papadimitriou C. H. and Steiglitz K. (1998),
{\it Combinatorial Optimization: Algorithm and Complexity}, 
Dover Publications, New York 
%%%(1998)%5

\bibitem{Parisi-pap} Parisi G. (1980), J. Phys A {bf 13} 1101 
%%%(1980)  

\bibitem{Martin-TSP} Percus A. and Martin O. C. (1996), 
Phys. Rev. Lett. {\bf 76}, 1188 
%%%(1996)

\bibitem{BC:Ray} Ray P., Chakrabarti B. K. and Chakrabarti A. (1989),
Phys. Rev. B {\bf{39}} 11828 
%%(1989)%13


\bibitem{Heiko-2} Rieger H. (2005), in Das and Chakrabarti (2005),
pp 69-97 
%%(2005b). 

\bibitem{Roland-Cerf} Roland J. and Cerf N. J. (2001), 
Phys. Rev. A {\bf 65} 042308

\bibitem{Laguna} Rodriguez-Laguna J. (2007), J. Stat. Mech.
P05008 
%%(pre-print in quant-ph/0702169).

\bibitem{Sachdev} S. Sachdev (1999),
 {\textit {Quantum Phase Transitions}}, Cambridge Univ. Press,
Cambridge 
%%(1999)

\bibitem{Santoro-Sc} Santoro G. E., Marto\u{n}\'{a}k R., Tosatti E. 
and Car R. (2002), Science, {\bf 295} 2427 
%%%(2002)

\bibitem{Santoro-rev} Santoro G. E. and 
Tosatti E. (2006), J. Phys. A {\bf 39} R393 
%%%%(2006)%20

\bibitem{Santoro-Nature} Santoro G. E. and Tosatti E. (2007),
{\it News and Views} in Nature Physics  {\bf 3}, 593 

%\bibitem[AD] A. Das, A. Dutta and B. K. Chakrabarti, cond-mat/0310381,
%B. K. Chakrabarti and A. Das cond-mat/0312611 \\

\bibitem{Sarandy-Lidar} Sarandy M. S., Wu L.-A., Lidar D. A. (2004),
Quantum Information Processing {\bf 3} 331

\bibitem{Sarjala} Sarjala M., Pet\"{a}j\"{a} V. and Aalva M. (2006), 
J. Stat. Mech. PO 1008 

\bibitem{Kris} Sengupta K., Powell S. and Sachdev S. (2004),
Phys. Rev. A {\bf 69} 053616

\bibitem{Kris-2}
Sengupta K., Sen D., and Mondal S. (2008)
arXiv:0710.1712 (to appear in Phys. Rev. Lett.)

\bibitem{Sherrington} Sherrington D. and Kirkpatrick S. (1975), 
Phys. Rev. Lett.
{\bf 35} 1792 
%%(1975)

\bibitem{Moshe}Schechter M. and Laflorencie N. (2006),
Phys. Rev. Lett. {\bf 97}, 137204

\bibitem{Silevitch}
Silevitch D. M., Bitko D., Brooke J., Ghosh S., Aeppli G., 
Rosenbaum T.F. (2007), Nature {\bf 448} 567

\bibitem{Somma}Somma R. D., Batista C. D., and Ortiz G. (2007),
Phys. Rev. Lett. {\bf 99} 030603

\bibitem{Sei} Suzuki S. and Okada M. (2005a), 
J. Phys. Soc. Japan {\bf 74} 1649 
%\bibitem{Sei-2} Suzuki S. and Okada M. (2005b), in Das and Chakrabarti (2005),
% pp 207-237
%{\textit{Quantumam Annealing and Related
%Optimization Methods}}, A. Das and B. K. Chakrabarti (Eds.),
%Lecture Note in Physics, {\bf 679},
%Springer-Verlag, Heildelberg (2005)%34
\bibitem{Hogg} Steffen M., Dam v. W., Hogg T., Breyta G., Chuang I. (2003),
Phys. Rev. Lett. {\bf 90} 067903 

\bibitem{Stella} Stella L., Santoro G., and Tosatti E. A. (2005),
Phys. Rev. B {\bf 72} 014303.

\bibitem{Thirumalai} Thirumalai D., Li Q. and Kirkpatrick T. R. (1989),
J. Phys. A {\bf 22} 2339 
%%(1989) 

\bibitem{BC:Wu} Wu W., Ellman B., Rosenbaum T. F.,
Aeppli G. and Reich D. H. (1991), Phys. Rev.
 Lett. {\bf{67}} 2076 
%%%(1991)

\bibitem{BC:Wu-2} Wu W., Bitko D., Rosenbaum T. F.
and Aeppli G. (1993a), Phys. Rev.
 Lett. {\bf{74}} 3041 
%%(1993a)%15

\bibitem{HR:qsg-exp1}
  Wu W., Bitko D., Rosenbaum T. F., and  Aeppli G. (1993b),
  Phys. Rev. Lett. {\bf 71} 1919 
%%%(1993b)

\bibitem{Ye} Ye J., Sachdev S. and Read N. (1993),
Phys. Rev. Lett. {\bf 70} 4011

\bibitem{zurek} Zurek W. H. (1985), Nature {\bf 317} 505

\end{thebibliography}
\end{document}